\documentclass{mn2e}
\usepackage{psfig}
\usepackage[authoryear]{natbib}

\newcommand\ros{{\it ROSAT}}
\newcommand\xmm{{\it XMM-Newton}}
\newcommand\chandra{{\it Chandra}}

\newcommand\nh{N_{H}}
\newcommand\ukdeep{13$^{\rm H}$}
\newcommand\pn{pn}
\newcommand\mos{MOS}

\title{X-ray spectra of sources in the \ukdeep\ {\em XMM-Newton} / {\em
Chandra} deep field}
\author[Page et al.]{
M.J. Page$^{1}$, 
N.S. Loaring$^{1}$, 
T. Dwelly$^{1}$,
K.O. Mason$^{1}$,
I. McHardy$^{2}$, 
\newauthor 
K. Gunn$^{2}$,
D. Moss$^{2}$, 
T. Sasseen$^{3}$,
F. Cordova$^{4}$,
J. Kennea$^{5}$,
N. Seymour$^{6}$
\\
$^{1}$Mullard Space Science Laboratory, University College London,
Holmbury St Mary, Dorking, Surrey, RH5 6NT, UK\\
$^{2}$Department of Physics and Astronomy, University of Soutampton, 
Southampton, SO17 1BJ, UK\\
$^{3}$Department of Physics, University of California, Santa Barbara, CA 93106, USA\\
$^{4}$University of California, Riverside, 900 University Avenue, Riverside, CA
92521, USA\\
$^{5}$Department of Astronomy and Astrophysics, 525 Davey Lab, Pennsylvania
State University, University Park, PA 16802, USA\\
$^{6}$Spitzer Science Center, California Institute of Technology, Mail Code
220-6, 1200 East California Boulevard, Pasadena, CA 91125, USA
}

\date{}

\begin{document}
\maketitle

\begin{abstract}
We present the X-ray spectra of 86 optically-identified sources in the \ukdeep\
\xmm/\chandra\ deep field which have $>70$~X-ray counts. The majority of these 
sources have 2-10 keV fluxes between $10^{-15}$ and
$5\times 10^{-14}$ erg~cm$^{-2}$~s$^{-1}$.  The sample consists
of 50 broad line AGN, 25 narrow emission line galaxies, 6 absorption line
galaxies, and 5 Galactic stars.  The majority (42/50) of the broad line AGN
have X-ray spectra which are consistent with a power law shape. They have a
mean photon index $\langle\Gamma\rangle=2.0\pm0.1$ and an intrinsic dispersion
$\sigma_{\Gamma}=0.4\pm 0.1$.  Three of the broad line AGN show curved spectra,
with more emission near the high and low energy ends of the spectrum relative
to the emission in the 1-2 keV range than can be reproduced by the power
law model.  Five BLAGN show a deficit of soft X-rays, indicating absorption.
We consider a source to be significantly absorbed if a power law model fit is
rejected with $>99\%$ confidence and an absorbed power law model produces an
acceptable fit, or if the best-fit power law is abnormally hard ($\Gamma < 1$).
Significant absorption is more common in the narrow emission line galaxies
(13/25) and absorption line galaxies (2/6) than in the broad line AGN (5/50),
but is not
universal in any of these classes of object. 
The majority of the 20 absorbed sources have X-ray spectra 
consistent
with a simple cold photoelectric absorber, but a significant minority (6/20)
require more complex models with either 
an additional component of soft X-ray emitting
plasma, or an ionised absorber.  Of the 16 narrow emission line and absorption
line galaxies which do not show evidence for X-ray absorption, only 2 objects
are likely to be powered by star formation, and both have 2--10 keV X-ray
luminosities of $\le 10^{40}$~erg~s$^{-1}$.  The X-ray emission in the other 14
unabsorbed NELGs and galaxies is most likely powered by AGN, which are 
not detected in the optical
because they are outshone by their luminous host galaxies.  The Galactic stars
show multi-temperature thermal spectra which peak between 0.5 and 1
keV. Star/AGN discrimination is possible for 4 of the 5 stars solely from
their X-ray spectra.

\end{abstract}

\section{Introduction}
\label{sec:introduction}
The cosmic X--ray background (CXRB) is widely interpreted as the 
integrated X-ray emission from many
discrete sources. Active galactic nuclei (AGN) are expected to contribute 
the majority of the CXRB, and
since the late 1980s the leading `synthesis models' have invoked a combination
of absorbed and unabsorbed AGN to reproduce the overall intensity and broadband
spectral shape of the CXRB.  Typically these synthesis models are based on AGN
`unification schemes' \citep{antonucci93} in which each AGN contains an
obscuring torus, and the orientation to our line of sight determines the
absorption \citep[e.g. ][]{setti89, comastri95, gilli99}.

The energy density of the CXRB peaks at $\sim 30$ keV, but the CXRB has only
been substantially resolved into sources at lower energies.  In the 0.5-2 keV
band, the deepest surveys with \ros\ resolved 70-80\% of the X--ray background
into individual objects, the majority of which are broad-line AGN
\citep[e.g. ][]{mchardy98,hasinger98}. Deep surveys performed with \xmm\ and
\chandra\ have increased the resolved fraction to 80-90\% in the 0.5-2~keV
band, and have also resolved a significant fraction ($>70\%$) in the 2-10 keV
band \citep[e.g. ][]{hasinger01,giacconi02,alexander03}. The broadband X-ray
spectral properties of the objects detected in the \xmm\ and \chandra\ surveys
support the hypothesis that the majority of the CXRB is due to AGN with a wide
range of photoelectric absorption.

However, there have been a number of findings that are contrary to the
assumptions and predictions of the synthesis models.  Firstly, the optical and
X-ray spectroscopic characteristics of the X-ray sources are sometimes
discrepant. There are now many examples of AGN which appear to show
photoelectric absorption from large column densities ($\ge10^{22}$~cm$^{-2}$)
of material, but which exhibit little attenuation to their optical and
ultraviolet radiation \citep[e.g. ][]{page01a,maiolino01}. On the other hand,
there are also examples of AGN for which the broad optical emission lines are
weak or absent, suggesting obscuration of the broad line region by dust, but
which show little or no absorption in their X-ray spectra
\citep[e.g. ][]{pappa01,barcons03,mateos05b}.

Secondly the absorbed sources, and those without broad optical emission lines
in their optical spectra (often referred to as `type 2' AGN), are typically
found at lower redshifts than predicted by the synthesis models
\citep{barger03,barger05,mainieri05}. This suggests that the absorption
properties of AGN have some dependence on either luminosity or redshift, in
contrast to the very simple geometric unification schemes employed in the
earlier versions of the synthesis models, which were independent of luminosity
or redshift. However, this issue is still controversial
\citep[e.g. see][]{treister04, dwelly05} and determining the absorption
characteristics of AGN as a function of redshift and luminosity is one of the
most important objectives of X-ray survey science.

The \ukdeep\ \xmm\ / \chandra\ deep field is a project to investigate the
astrophysics of the major contributors to the CXRB, particularly sources around
the break in the X--ray source counts where the contribution to the CXRB per
log-flux interval peaks. In order to understand the phenomena, processes and
conditions in these sources we combine the high quality X--ray spectra of \xmm\
with the precise positions of \chandra\ and a host of other multi-wavelength
data, including a deep 1.4 GHz survey from the VLA \citep{seymour04}. 
The \ukdeep\ field is centred at $13^{h}34^{m}37^{s} +
37^{\circ}54^{'}44^{''}$, was the 
location of one of the deepest {\em ROSAT} surveys
\citep{mchardy98} and is a region of extremely low Galactic absorption (${\rm
N_{H}} \sim 7 \times 10^{19}$~cm$^{-2}$).

In this paper we present the European Photon Imaging Camera 
\citep[EPIC, ][]{turner01,struder01} spectra of the \ukdeep\ sources that have
been optically identified and which have sufficient X-ray counts to yield a
meaningful X-ray spectrum. At present, this amounts to 86 sources.  Throughout
this paper we assume $H_{0}=70$~km~s$^{-1}$~Mpc$^{-1}$, $\Omega_{\Lambda}=0.7$,
and $\Omega_{m}=0.3$.  We define a power law spectrum as $dN/dE = AE^{-\Gamma}$
where $N$ is the number of photons, $E$ is photon energy, $\Gamma$ is the
photon index and $A$ is the normalisation. The source numbering scheme used
here is identical to that used in \citet{loaring05}.

\section{X-ray Observations and data reduction}
The \ukdeep\ field was observed three times with \xmm\ for a total exposure
time of 200\,ks. When deadtime and periods of high background are excluded, the
total live exposure time is 120\,ks for the \pn\ camera and 130\,ks for each of
the
\mos\ cameras.  
The \xmm\ observations are fully described in \citet{loaring05}.

The EPIC spectra were produced from the same event lists, filtered with the
same good time intervals, and using the same version of the \xmm\ Science
Analysis System (SAS 6.0) as the 
source lists presented in
\citet{loaring05}.  Spectra were produced for all optically-identified sources
with $>70$ source counts.  Source counts were accumulated in elliptical regions
centred on the positions reported by \citet{loaring05}. The size and
ellipticity of each source extraction region was determined by the off-axis
angle of the X-ray source, and its countrate. The point spread function is
relatively symmetrical near the centre of the \xmm\ field of view but becomes
elongated in the tangential direction at larger off-axis angles.  Therefore to
optimise signal to noise the
ellipticity of the source extraction regions increases with off-axis angle,
from 0 at the optical axis to 2 near the edge of the field.  The minor axes of
our source regions vary from 10$\arcsec$ to 20$\arcsec$ depending on the 
overall
source countrates. There are several pairs of 
sources which are close enough
together that the nominal extraction regions would overlap. In such cases the
two regions were renormalised so that they do not overlap, with the ratio of
the minor axes set to the ratio of the countrates of the two sources.

All valid event patterns (PATTERN 0--12) were used in constructing the \mos\
spectra. For the pn spectra, single and double events (PATTERN 0--4) were used
for channel energies $>0.4$~keV, and only single events were used for lower
energy channels.  Channels close to the energies of the strongest instrumental
emission lines (Cu-K at 7.8 keV in pn and Al-K at 1.7 keV in \mos) were
excluded from the spectra.  Response matrices and effective area files were
constructed for each source region using the SAS tasks {\small RMFGEN} and
{\small ARFGEN} respectively.  For each source the different EPIC spectra were
combined to form a single spectrum over the 0.2--12 keV range, 
using the method of
\citet{pagedavissalvi03}. Spectra were grouped to a minimum of 30 counts per 
bin.

\section{Optical spectroscopic identification}
Optical spectroscopic observations were carried out in 2002 and 2003
at the William Herschel Telescope, 
using the AF2/WYFFOS multi-fibre positioner
and spectrograph, and at the W.M. Keck observatory, using the LRIS and DEIMOS
multi-object spectrographs. At present, 115 \xmm\ and \chandra\ sources have
optical spectroscopic identifications and redshifts. Full details of the 
observations and the optical
spectra will be presented in Loaring et~al. (in prep).

\begin{table}
\caption{Power law fits to the \xmm\ spectra. $A$ is the power law
normalisation at 1 keV in units of 
$10^{-6}$~photons~cm$^{-2}$~s$^{-1}$~keV$^{-1}$. $z$
is the redshift of the source. P is the null hypothesis probability
corresponding to $\chi^{2}/\nu$.}
\label{tab:pofits}
 \begin{tabular}{l@{\hspace{1.0mm}}cccc@{\hspace{2.0mm}}l}
 &&&&&\\
\hline
 Source&$z$&$\Gamma$&$A$&$\chi^{2}/\nu$&P\\
 &&&&&\\
\hline
\multicolumn{5}{l}{\bf BLAGN:}&\\
&&&&&\\
5  &3.43&$1.89^{+0.63}_{-0.48}$&$2.4^{+0.7}_{-0.7}$&1 / 2&0.67\\
7  &1.68&$2.60^{+0.47}_{-0.39}$&$1.0^{+0.3}_{-0.3}$&10 / 9&0.36\\
8  &0.92&$1.75^{+0.66}_{-0.47}$&$2.5^{+0.8}_{-0.8}$&1 / 2&0.66\\
12 &2.11&$1.73^{+0.38}_{-0.36}$&$1.1^{+0.2}_{-0.2}$&8 / 9&0.52\\
13 &1.08&$2.22^{+0.15}_{-0.13}$&$13.6^{+1.1}_{-1.1}$&33 / 23&$7.3\times 10^{-2}$\\
19 &0.97&$2.46^{+0.19}_{-0.17}$&$3.1^{+0.3}_{-0.3}$&18 / 17&0.38\\
24 &0.73&$2.45^{+0.27}_{-0.25}$&$1.8^{+0.3}_{-0.3}$&10 / 15&0.82\\
27 &2.81&$1.81^{+0.32}_{-0.29}$&$1.0^{+0.2}_{-0.2}$&12 / 9&0.20\\
30 &2.14&$1.71^{+0.10}_{-0.10}$&$4.8^{+0.3}_{-0.3}$&57 / 36&$1.5\times 10^{-2}$\\
32 &1.61&$1.92^{+0.08}_{-0.08}$&$8.8^{+0.4}_{-0.5}$&45 / 55&0.83\\
39 &1.64&$1.68^{+0.11}_{-0.11}$&$4.1^{+0.3}_{-0.3}$&39 / 35&0.28\\
40 &1.37&$0.21^{+0.51}_{-0.59}$&$0.14^{+0.10}_{-0.09}$&15 / 4&$5.6\times 10^{-3}$\\
41 &1.60&$1.24^{+0.24}_{-0.24}$&$1.0^{+0.2}_{-0.2}$&12 / 9&0.22\\
42 &1.34&$1.95^{+0.16}_{-0.16}$&$2.1^{+0.2}_{-0.2}$&32 / 18&$2.0\times 10^{-2}$\\
46 &2.42&$1.84^{+0.18}_{-0.18}$&$1.7^{+0.2}_{-1.5}$&19 / 17&0.31\\
48 &1.69&$1.78^{+0.19}_{-0.17}$&$1.4^{+0.2}_{-0.2}$&8 / 12&0.82\\
50 &0.85&$1.63^{+0.07}_{-0.08}$&$5.2^{+0.3}_{-0.3}$&62 / 45&$4.7\times 10^{-2}$\\
52 &0.39&$3.36^{+0.18}_{-0.17}$&$2.6^{+0.3}_{-0.3}$&60 / 30&$1.0\times 10^{-3}$\\
53 &0.77&$2.62^{+0.25}_{-0.22}$&$2.0^{+0.3}_{-0.3}$&30 / 17&$2.9\times 10^{-2}$\\
60 &2.47&$1.94^{+0.32}_{-0.29}$&$0.90^{+0.16}_{-0.17}$&8 / 9&0.56\\
65 &1.14&$2.29^{+0.06}_{-0.06}$&$15.9^{+0.5}_{-0.5}$&153 / 98&$3.2\times 10^{-4}$\\
69 &1.19&$1.93^{+0.23}_{-0.21}$&$1.4^{+0.2}_{-0.2}$&10 / 14&0.73\\
73 &0.55&$1.71^{+0.35}_{-0.32}$&$1.4^{+0.3}_{-0.3}$&2 / 5&0.81\\
77 &2.59&$2.05^{+0.25}_{-0.24}$&$2.0^{+0.3}_{-0.3}$&16 / 13&0.26\\
80 &1.00&$1.47^{+0.48}_{-0.44}$&$0.71^{+0.22}_{-0.22}$&8 / 5&0.13\\
81 &1.57&$2.12^{+0.18}_{-0.16}$&$2.9^{+0.3}_{-0.3}$&23 / 21&0.36\\
93 &1.87&$2.14^{+0.42}_{-0.39}$&$0.55^{+0.12}_{-0.12}$&8 / 8&0.44\\
96 &1.50&$0.99^{+0.16}_{-0.18}$&$0.85^{+0.16}_{-0.16}$&31 / 11&$1.3\times 10^{-3}$\\
97 &1.36&$1.78^{+0.06}_{-0.06}$&$10.0^{+0.5}_{-0.5}$&72 / 62&0.19\\
101&0.52&$1.61^{+0.21}_{-0.19}$&$1.8^{+0.2}_{-0.2}$&47 / 20&$6.3\times 10^{-4}$\\
108&2.07&$1.59^{+0.40}_{-0.38}$&$0.42^{+0.12}_{-0.12}$&2 / 3&0.52\\
113&0.57&$2.08^{+0.08}_{-0.07}$&$7.4^{+0.3}_{-0.3}$&87 / 56&$5.3\times 10^{-3}$\\
119&0.26&$1.76^{+0.08}_{-0.08}$&$6.3^{+0.4}_{-0.4}$&40 / 50&0.79\\
121&1.14&$2.80^{+0.10}_{-0.10}$&$7.1^{+0.4}_{-0.4}$&57 / 56&0.46\\
129&1.89&$2.25^{+0.19}_{-0.18}$&$1.8^{+0.2}_{-0.2}$&15 / 18&0.65\\
132&0.71&$1.10^{+0.08}_{-0.09}$&$4.4^{+0.4}_{-0.4}$&115 / 31&$1.4\times 10^{-11}$\\
134&1.18&$1.22^{+0.07}_{-0.08}$&$4.0^{+0.3}_{-0.3}$&73 / 39&$7.5\times 10^{-4}$\\
143&1.89&$1.95^{+0.10}_{-0.09}$&$5.9^{+0.3}_{-0.3}$&46 / 45&0.45\\
161&2.01&$2.18^{+0.42}_{-0.37}$&$0.99^{+0.23}_{-0.23}$&3 / 7&0.88\\
162&1.77&$1.99^{+0.28}_{-0.25}$&$1.0^{+0.2}_{-0.2}$&13 / 10&0.25\\
164&1.49&$2.34^{+0.77}_{-0.68}$&$0.39^{+0.13}_{-0.15}$&8 / 5&0.18\\
173&2.48&$2.13^{+0.27}_{-0.26}$&$1.3^{+0.2}_{-0.2}$&14 / 11&0.25\\
178&2.66&$1.75^{+0.50}_{-0.45}$&$0.45^{+0.12}_{-0.14}$&3 / 4&0.62\\
190&2.82&$1.93^{+0.20}_{-0.20}$&$2.3^{+0.3}_{-0.3}$&15 / 15&0.42\\
196&1.38&$1.53^{+0.21}_{-0.21}$&$2.0^{+0.3}_{-0.3}$&11 / 10&0.33\\
208&1.39&$2.01^{+0.16}_{-0.16}$&$4.7^{+0.4}_{-0.4}$&17 / 23&0.79\\
212&1.63&$2.04^{+0.11}_{-0.10}$&$8.7^{+0.6}_{-0.6}$&32 / 35&0.63\\
217&2.28&$1.61^{+0.20}_{-0.20}$&$2.6^{+0.4}_{-0.4}$&16 / 13&0.23\\
218&1.89&$1.84^{+0.20}_{-0.19}$&$3.6^{+0.4}_{-0.4}$&30 / 17&$2.7\times 10^{-2}$\\
221&1.62&$1.85^{+0.12}_{-0.12}$&$9.1^{+0.7}_{-0.7}$&34 / 33&0.40\\

 \end{tabular}
 \end{table}
 \begin{table}
 \begin{tabular}{l@{\hspace{1.0mm}}c@{\hspace{2.0mm}}c@{\hspace{1.2mm}}c@{\hspace{1.2mm}}c@{\hspace{2.0mm}}l}
 \multicolumn{6}{l}{{\bf Table \ref{tab:pofits}.} continued\vspace{2mm}}\\
 &&&&&\\
\hline
 Source&$z$&$\Gamma$&$A$&$\chi^{2}/\nu$&P\\
 &&&&&\\
\hline
\multicolumn{5}{l}{\bf NELGs:}&\\
&&&&&\\
1  &0.39&$2.09^{+0.57}_{-0.49}$&$2.5^{+0.7}_{-0.7}$&1 / 1&0.48\\
20 &0.17&$2.53^{+0.31}_{-0.27}$&$1.3^{+0.2}_{-0.2}$&4 / 7&0.77\\
56 &0.06&$2.54^{+0.58}_{-0.49}$&$0.46^{+0.17}_{-0.18}$&12 / 4&$1.7\times 10^{-2}$\\
62 &0.43&$0.60^{+0.99}_{-1.00}$&$0.18^{+0.15}_{-0.13}$&5 / 3&0.18\\
66 &0.29&$0.64^{+0.61}_{-0.44}$&$0.26^{+0.15}_{-0.14}$&1 / 3&0.77\\
67 &0.34&$0.70^{+0.09}_{-0.10}$&$1.8^{+0.2}_{-0.2}$&111 / 30&$3.0\times 10^{-11}$\\
76 &0.49&$-0.12^{+0.57}_{-0.75}$&$0.11^{+0.10}_{-0.08}$&3 / 4&0.60\\
94 &0.85&$1.45^{+0.53}_{-0.52}$&$0.21^{+0.09}_{-0.09}$&6 / 3&$9.8\times 10^{-2}$\\
100&0.27&$0.52^{+0.72}_{-0.75}$&$0.28^{+0.17}_{-0.15}$&14 / 3&$3.6\times 10^{-3}$\\
103&0.55&$1.51^{+0.15}_{-0.14}$&$1.4^{+0.2}_{-0.2}$&29 / 18&$4.3\times 10^{-2}$\\
109&0.57&$0.59^{+0.09}_{-0.10}$&$1.5^{+0.2}_{-0.2}$&87 / 24&$4.7\times 10^{-9}$\\
115&0.23&$1.85^{+0.14}_{-0.13}$&$4.6^{+0.4}_{-0.4}$&17 / 25&0.88\\
118&0.06&$1.95^{+0.49}_{-0.44}$&$0.37^{+0.11}_{-0.11}$&3 / 5&0.76\\
126&1.09&$0.42^{+0.68}_{-0.70}$&$0.13^{+0.11}_{-0.07}$&7 / 3&$7.2\times 10^{-2}$\\
131&0.22&$2.03^{+0.51}_{-0.45}$&$0.37^{+0.11}_{-0.12}$&4 / 4&0.36\\
142&0.31&$0.69^{+0.11}_{-0.11}$&$1.4^{+0.2}_{-0.2}$&101 / 23&$9.2\times 10^{-12}$\\
153&0.43&$1.87^{+0.57}_{-0.52}$&$0.33^{+0.12}_{-0.12}$&4 / 3&0.22\\
165&0.56&$0.92^{+0.16}_{-0.17}$&$1.00^{+0.2}_{-0.2}$&26 / 13&$1.6\times 10^{-2}$\\
175&0.36&$1.66^{+0.12}_{-0.12}$&$4.6^{+0.4}_{-0.4}$&68 / 34&$5.0\times 10^{-4}$\\
186&0.81&$1.60^{+0.32}_{-0.30}$&$0.96^{+0.18}_{-0.18}$&1 / 5&0.94\\
193&0.12&$2.11^{+0.62}_{-0.49}$&$0.63^{+0.20}_{-0.20}$&2 / 4&0.69\\
200&0.88&$1.36^{+0.35}_{-0.34}$&$0.80^{+0.20}_{-0.20}$&5 / 7&0.62\\
213&0.80&$0.15^{+0.22}_{-0.24}$&$1.0^{+0.3}_{-0.3}$&16 / 7&$2.4\times 10^{-2}$\\
222&0.26&$0.91^{+0.22}_{-0.22}$&$2.3^{+0.5}_{-0.5}$&17 / 10&$7.4\times 10^{-2}$\\
225&0.26&$2.08^{+0.41}_{-0.36}$&$1.8^{+0.5}_{-0.5}$&17 / 4&$2.0\times 10^{-3}$\\

\hline
\multicolumn{5}{l}{\bf galaxies:}&\\
&&&&&\\
37 &0.26&$2.19^{+0.13}_{-0.12}$&$3.7^{+0.3}_{-0.3}$&16 / 18&0.62\\
55 &1.08&$2.10^{+0.28}_{-0.25}$&$1.0^{+0.2}_{-0.2}$&4 / 10&0.94\\
92 &0.29&$0.56^{+0.18}_{-0.18}$&$0.51^{+0.12}_{-0.12}$&37 / 12&$2.6\times 10^{-4}$\\
98 &0.36&$-0.24^{+0.49}_{-0.58}$&$0.08^{+0.07}_{-0.05}$&12 / 5&$3.5\times 10^{-2}$\\
117&0.62&$2.02^{+0.67}_{-0.64}$&$0.57^{+0.21}_{-0.21}$&9 / 4&$6.0\times 10^{-2}$\\
184&0.55&$1.55^{+0.31}_{-0.28}$&$1.8^{+0.3}_{-0.3}$&14 / 10&0.16\\

\hline
\multicolumn{5}{l}{\bf stars:}&\\
&&&&&\\
49 &$-$&$2.09^{+0.16}_{-0.16}$&$2.6^{+0.3}_{-0.3}$&37 / 19&$7.1\times 10^{-3}$\\
61 &$-$&$3.09^{+0.18}_{-0.16}$&$1.3^{+0.2}_{-0.2}$&110 / 19&$6.8\times 10^{-15}$\\
120&$-$&$2.45^{+0.26}_{-0.25}$&$1.9^{+0.3}_{-0.4}$&31 / 11&$1.1\times 10^{-3}$\\
140&$-$&$2.56^{+0.01}_{-0.01}$&$163^{+2}_{-2}$&8027 / 165&\ $< 10^{-30}$\\
179&$-$&$1.96^{+0.81}_{-0.64}$&$0.53^{+0.20}_{-0.20}$&2 / 4&0.70\\
 \end{tabular}
\end{table}

\section{Results}

\begin{figure*}
\begin{center}
\leavevmode
\psfig{figure=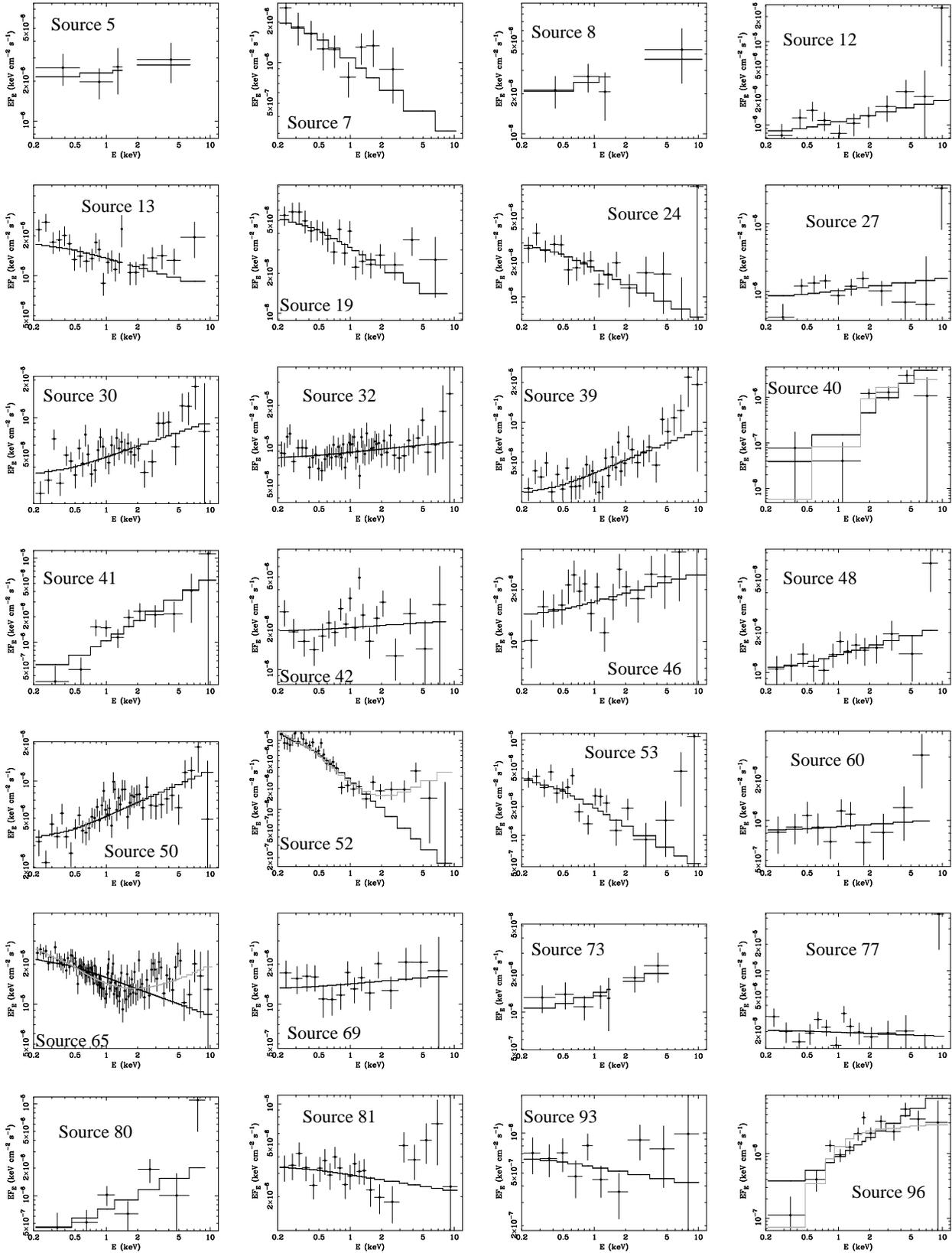,width=165truemm}
\caption{X-ray spectra of the BLAGN (data points) together with power
law models (black stepped lines). All spectra are shown in the observer frame. 
Both model and data have been divided by the
product of the effective area and Galactic transmission 
as a function of energy. For those objects which required a more
complex spectral model, this is shown as a grey stepped line. For sources 52,
65 and 113 the grey line is the two power law model. For sources 40, 96 and
134, the grey line is a power law with cold absorption intrinsic to the AGN.
For source 101 the grey line is a power law with an ionised absorber,
and for source
132 the grey line is an absorbed power law with an optically thin thermal
plasma contributing at soft energies.}
\label{fig:blagnspecs}
\end{center}
\end{figure*}
\begin{figure*}
\begin{center}
\leavevmode
\psfig{figure=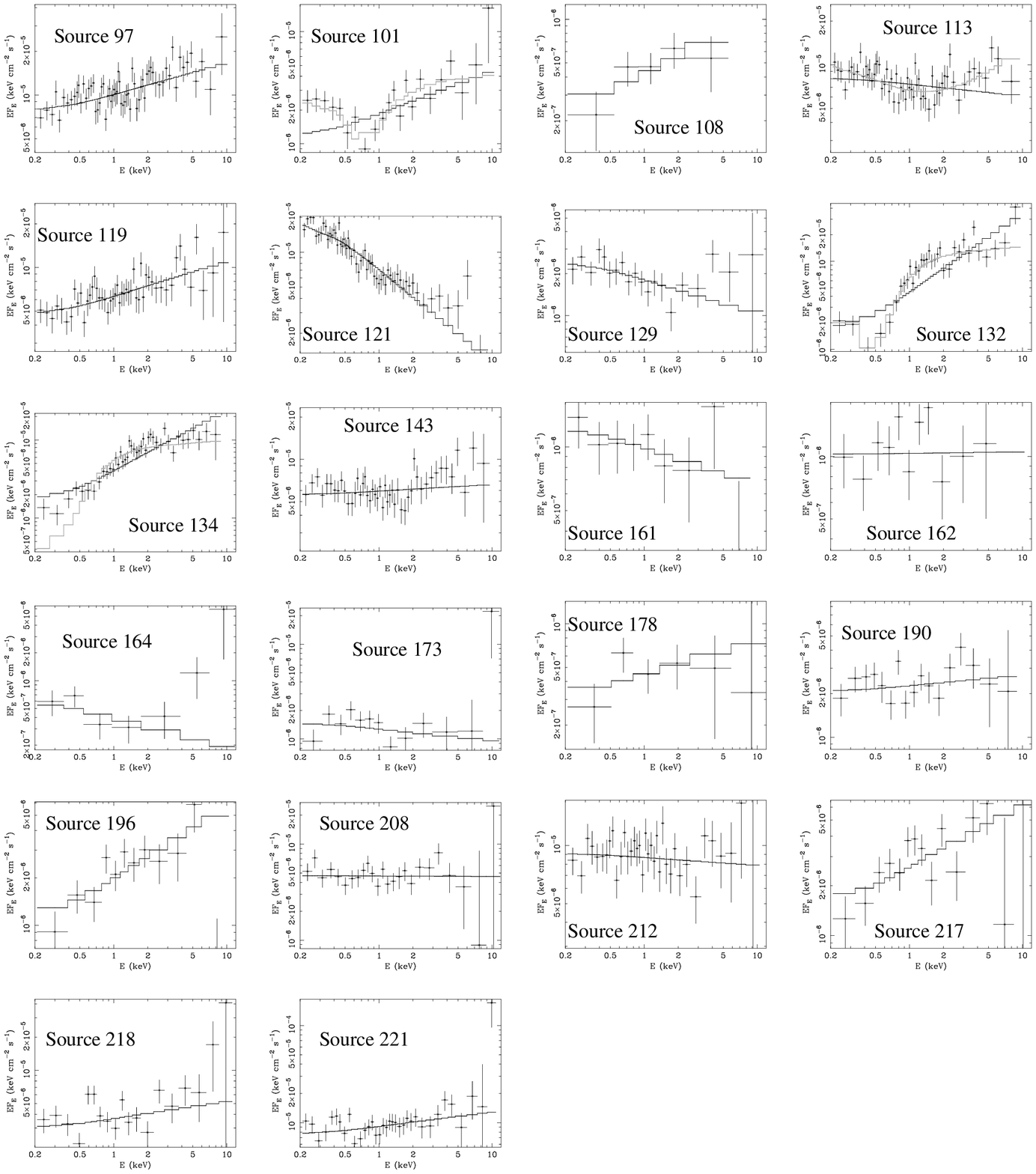,width=165truemm}
\end{center}
{\bf Figure \ref{fig:blagnspecs}} {\it X-ray spectra of the BLAGN continued}
\end{figure*}

\begin{figure*}
\begin{center}
\leavevmode
\psfig{figure=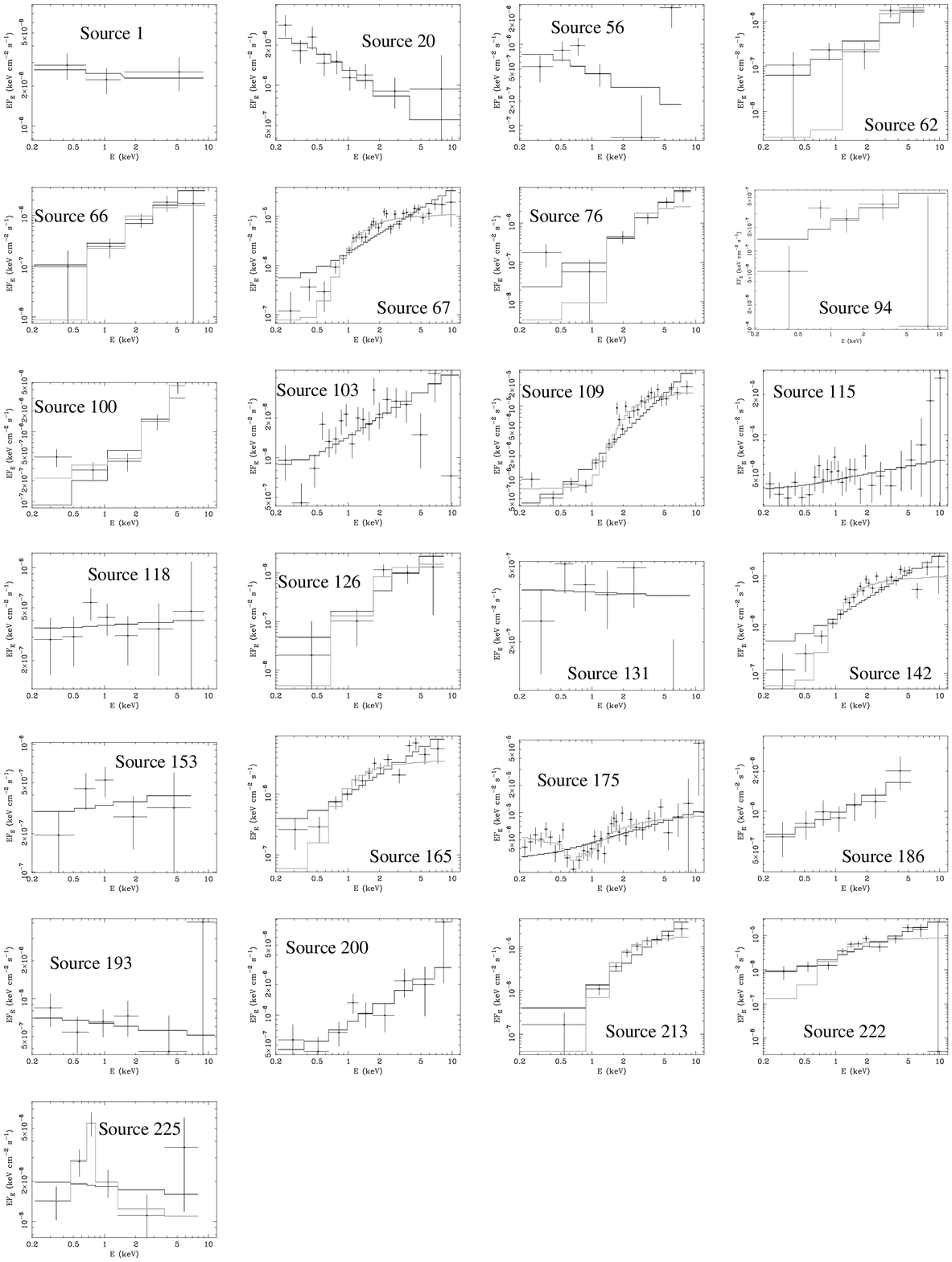,width=165truemm}
\caption{X-ray spectra of the NELGs (data points) together with power
law models (black stepped lines). All spectra are shown in the observer frame.
Both model and data have been divided by the
product of effective area and Galactic transmission 
as a function of energy. For those sources which were fitted
with more complex models, the best fit is shown as a grey
stepped line. For sources 62, 66, 67, 76, 126, 142, 165, 213, and 222 the grey
line is a power law model with cold absorption. For source 175 the grey line
shows a power law with an ionised absorber, and for source 109 the grey line is an
absorbed power law with a second, unabsorbed, powerlaw component. For sources
100 and 225 the grey line shows an absorbed power law with an optically thin thermal
plasma contributing at soft energies.
 }
\label{fig:nelgspecs}
\end{center}
\end{figure*}

There are 183 sources with $>70$ source counts in the \ukdeep\ field, 86 of
which are optically identified. The majority of these 
identified sources have 2-10 keV fluxes in the range 
$10^{-15}$ -- $5\times 10^{-14}$ erg~cm$^{-2}$~s$^{-1}$. 
We have divided these 86 sources into four categories according to
their optical spectroscopic properties: 50 
broad line AGN (BLAGN) which have one
or more emission lines with full-width half maximum (FWHM) $>
1000$~km~s$^{-1}$, 25 narrow emission line galaxies (NELGs) which have emission
lines of FWHM $< 1000$~km~s$^{-1}$, 6 absorption line galaxies which do not
exhibit any emission lines in our optical spectra (hereafter simply
`galaxies'), and 5 Galactic stars.  

\begin{figure}
\begin{center}
\leavevmode
\psfig{figure=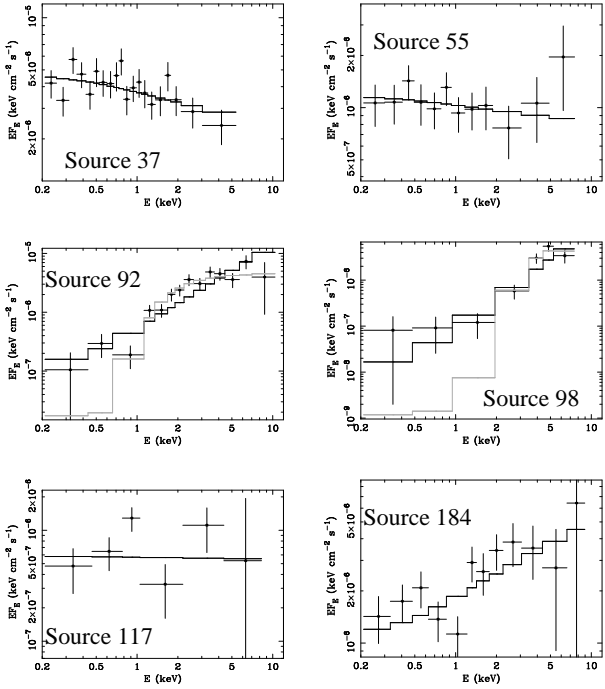,width=80truemm}
\caption{X-ray spectra of the galaxies (data points) together with power
law models (black stepped lines). All spectra are shown in the observer frame.
Both model and data have been divided by the product of 
effective area and Galactic transmission 
as a function of energy. For sources 92 and 98 a better fitting
model of a power law
absorbed
by cold material is also shown, as a grey stepped line.}
\label{fig:galaxyspecs}
\end{center}
\end{figure}

\begin{figure}
\begin{center}
\leavevmode
\psfig{figure=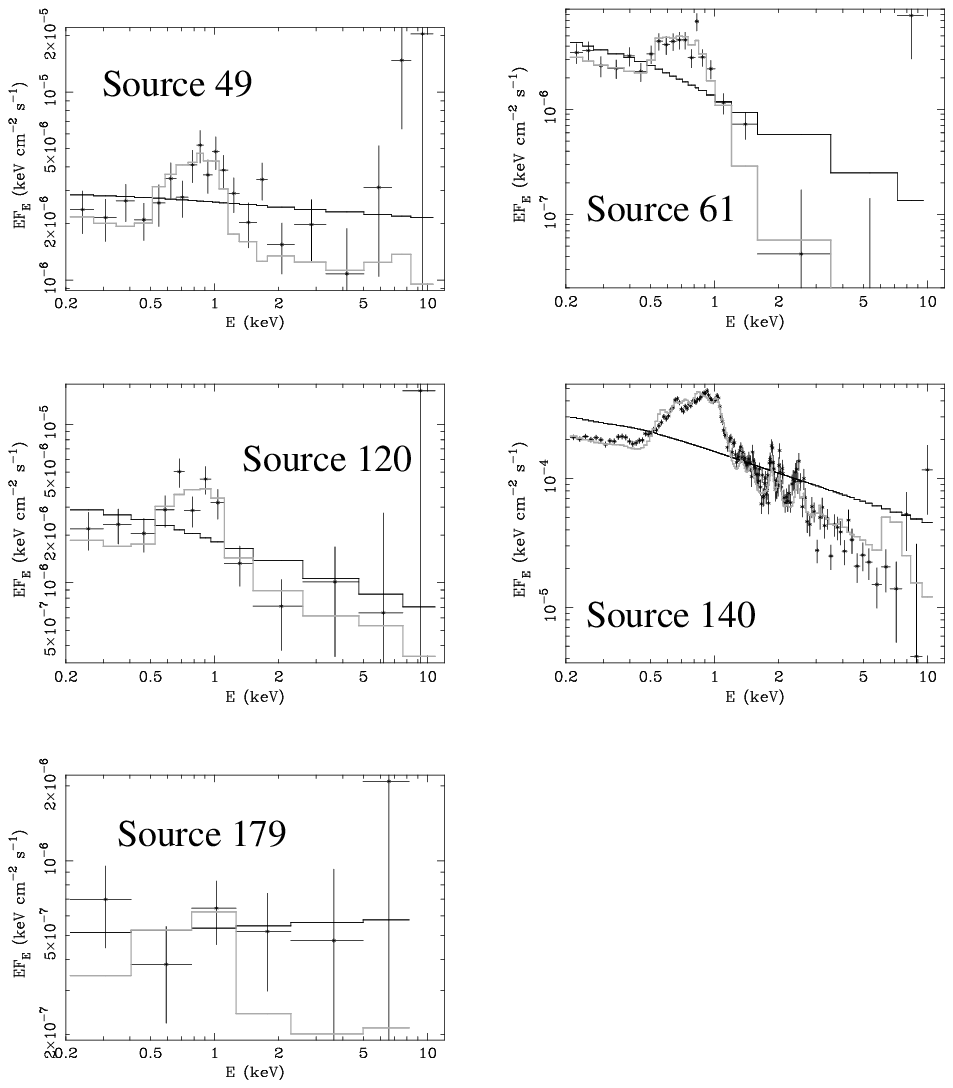,width=80truemm}
\caption{X-ray spectra of the Galactic stars (data points) together with power
law models (black stepped lines) and multi-temperature optically-thin thermal
models (grey stepped lines). All spectra are shown in the observer frame. 
Both models and data have been divided by the product of 
effective area and Galactic transmission as a function of energy.}
\label{fig:starspecs}
\end{center}
\end{figure}

As a starting point
for the X-ray spectral investigations we fitted each spectrum with a power law
model, and fixed Galactic absorption ($\nh=7\times 10^{-19}$~cm$^{-2}$). The
results of these fits are given in Table \ref{tab:pofits}. The individual
spectra, grouped by optical counterpart type, are shown along with the best-fit
power law models in figs. \ref{fig:blagnspecs}, \ref{fig:nelgspecs},
\ref{fig:galaxyspecs}, and \ref{fig:starspecs}. In order to show the intrinsic
shapes of the spectra, both the spectra and the models in figs.
\ref{fig:blagnspecs}, \ref{fig:nelgspecs}, \ref{fig:galaxyspecs}, and
\ref{fig:starspecs} have been divided by the product of the EPIC effective area
and the Galactic transmission on a channel by channel basis. For the sample
size of 86 spectra, we consider a null hypothesis probability of 1\% to be an
appropriate threshold between acceptable and unacceptable goodness of fit. We
now describe the EPIC spectral fits for each of the four optical spectroscopic
subclasses.

\subsection{BLAGN}

The BLAGN represent the largest subsample, numbering 50 sources. The power law
model is quite successful for these sources: 42 can be fitted acceptably with the
power law model, and 8 cannot. The mean best fit spectral index of the
acceptably fitted sources is $\langle \Gamma
\rangle = 1.90\pm 0.06$.

Inspection of Fig. \ref{fig:blagnspecs} reveals that the poorly fitting spectra
deviate from a power-law shape in several different ways. Just over half of the
poor-fitting spectra are relatively hard, (sources 40, 96, 101, 132 and 134),
while the other three are relatively soft (sources 52, 65 and 113).  The hard
spectrum sources could deviate from power law shapes because they are
photoelectrically absorbed. Therefore, we refitted these spectra, including
photoelectric absorption from cold gas at the same redshift as the AGN.  To
minimise the number of free parameters in the fit, we fixed the power law
slopes at a `typical' BLAGN value of $\Gamma=1.9$ and assumed Galactic
abundances \citep{anders89}.  The results are given in Table
\ref{tab:nhfits}. For three of the sources acceptable fits were obtained, while
for source 132 the fit was significantly better than the simple power law fit,
though still rejected at 99\% confidence. However, for source 101 there was no
improvement to the $\chi^{2}$ compared to the power law model, and the best fit
column density is zero.  For source 132, the largest contribution to $\chi^{2}$
comes from the two lowest-energy channels, which are in excess of the model,
while for source 101 there is a minimum around 0.7 keV (observed frame) which
the model cannot reproduce.

\begin{table}
\caption{Power law fits including cold photoelectric absorption and with fixed
$\Gamma = 1.9$. $A$ is the power law normalisation in units of
$10^{-6}$~photons~cm$^{-2}$~s$^{-1}$~keV$^{-1}$. Sources marked with an `*'
have acceptable $\chi^{2}$ with a simple powerlaw fit, but have been fitted with
the model including cold absorption because they have best-fit $\Gamma<1$.
Parameters and uncertainties marked with $\dag$ are constrained by the maximum
or minimum of the allowed range. P is the null hypothesis probability
corresponding to $\chi^{2}/\nu$.}
\label{tab:nhfits}
 \begin{tabular}{l@{\hspace{2.0mm}}cccl}
 &&&&\\
\hline
Source&$A$&$N_{H}$&$\chi^{2}/\nu$&P\\
&&($10^{22}$~cm$^{-2}$)&&\\
\hline
\multicolumn{4}{l}{\bf BLAGN:}&\\
&&&&\\
40&$  2.2^{+ 1.4}_{ -1.0}$&$ 19^{+ 20}_{ -9}$&  5 /  4&$0.34$\\
96&$  2.3^{+ 0.6}_{ -0.5}$&$  2.5^{+  1.7}_{ -1.0}$&  18 / 11&$7.6\times 10^{-2}$\\
101&$  1.8^{+ 0.2}_{ -0.2}$&$  0.00\dag^{+  0.04}_{ -0.00\dag}$&  54 / 20&$5.6\times 10^{-5}$\\
132&$ 11.3^{+ 1.2}_{ -1.1}$&$  0.83^{+  0.19}_{ -0.16}$&  53 / 31&$8.2\times 10^{-3}$\\
134&$  7.9^{+ 0.8}_{ -0.8}$&$  1.00^{+  0.30}_{ -0.24}$&  58 / 39&$2.4\times 10^{-2}$\\
\hline
\multicolumn{4}{l}{\bf NELGs:}&\\
&&&&\\
62$^{*}$&$  2.1^{+ 2.1}_{ -1.9}$&$  9.5^{+ 19.6}_{ -9.4}$&  7 /  3&$7.2\times 10^{-2}$\\
66$^{*}$&$  1.3^{+ 1.2}_{ -0.7}$&$  1.7^{+  3.9}_{ -1.2}$&  1 /  3&0.70\\
67      &$  8.7^{+ 1.1}_{ -1.0}$&$  1.17^{+  0.25}_{ -0.20}$& 38 / 30&0.15\\
76$^{*}$&$  2.4^{+ 9.7}_{ -1.2}$&$  9.7^{+ 61.8}_{ -6.3}$&  8 /  4&0.11\\
109      &$ 11.5^{+ 1.6}_{ -1.5}$&$  4.0^{+  0.9}_{ -0.7}$& 97 / 24&$8.6\times 10^{-11}$\\
126$^{*}$&$  1.4^{+ 1.0}_{ -0.7}$&$  10^{+ 14}_{ -6}$&  2 /  3&0.63\\
142      &$  7.7^{+ 1.1}_{ -1.0}$&$  1.4^{+  0.3}_{ -0.3}$& 31 / 23&0.13\\
165$^{*}$&$  2.9^{+ 0.6}_{ -0.6}$&$  0.96^{+  0.48}_{ -0.33}$& 22 / 13&$5.1\times 10^{-2}$\\
175      &$  4.7^{+ 0.6}_{ -0.4}$&$  0.01^{+  0.03}_{ -0.01\dag}$& 82 / 34&$8.5\times 10^{-6}$\\
184      &$  0.4^{+ 0.3}_{ -0.1}$&$  0.01^{+  0.26}_{ -0.01\dag}$& 25 /  3&$1.4\times 10^{-5}$\\
213$^{*}$&$ 14.3^{+ 4.4}_{ -3.7}$&$  8.6^{+  4.1}_{ -2.6}$&  5 /  7&0.62\\
222$^{*}$&$  6.9^{+ 1.9}_{ -1.8}$&$  0.64^{+  0.34}_{ -0.27}$& 21 / 10&$2.0\times 10^{-2}$\\
225      &$  1.9^{+ 0.6}_{ -0.5}$&$  0.00\dag^{+  0.06}_{  0.00\dag}$& 18 / 4&$1.3\times 10^{-3}$\\
\hline
\multicolumn{4}{l}{\bf galaxies:}&\\
&&&&\\
92      &$  3.7^{+ 0.9}_{ -0.8}$&$  2.1^{+  0.8}_{ -0.6}$& 15 / 12&0.26\\
98$^{*}$&$  5.9^{+ 4.7}_{ -2.6}$&$ 25^{+ 22}_{-12}$&  7 / 5&0.20\\
\end{tabular}
\end{table}

There are several reasonable modifications to the absorption model that might
better reproduce the spectra of sources 101 and 132.  
Firstly, there might be some extra
emission component at the softest energies. We tried two variants of such a
model: one including an emission line component, modelled as a low temperature
thermal plasma (mekal in {\small XSPEC}), and one including a component of
unabsorbed power law emission. Soft X-ray emission lines are prominent in the
spectra of nearby Seyfert 2 galaxies \citep[e.g.][]{sako01,kinkhabwala02}, and
although they are photoionised rather than collisionally ionised, the mekal
model is a reasonable approximation at the resolution and signal to noise ratio
of our EPIC spectra.  On the other hand,  
an unabsorbed power law component could
represent primary emission that has been scattered into our line of sight,
bypassing the absorbing medium. Such a model has been used widely
\citep[e.g.][]{turner97,franceschini03,caccianiga04} and has a simple and
convenient functional form.  The results for these two fits are given in Tables
\ref{tab:popofits} and \ref{tab:mekalfits}. The model with an unabsorbed power
law component fits the spectrum of source 132, but not the spectrum of source
101. The model including a soft thermal component provides a good fit to the
spectra of both objects.

\begin{table*}
\caption{Absorbed power law fits with an additional unabsorbed power law
component. $A_{1}$ is the normalisation of the primary, absorbed power law and
$A_{2}$ is the normalisation of the secondary, unabsorbed power law, both in
units of $10^{-6}$~photons~cm$^{-2}$~s$^{-1}$~keV$^{-1}$.  Parameters and
uncertainties marked with $\dag$ are constrained by the limits of the fit
range. P is the null hypothesis probability corresponding to $\chi^{2}/\nu$.}
\label{tab:popofits}
 \begin{tabular}{lccccl}
 &&&&&\\
\hline
 Source&$A_{1}$&$A_{2}$&$N_{H}$&$\chi^{2}/\nu$&P\\
 &&&($10^{22}$~cm$^{-2}$)&&\\
\hline
{\bf BLAGN:}&&&&&\\
&&&&&\\
 101&$ 1.7^{+ 1.4}_{-0.9}$&$1.6^{+0.3}_{-0.3}$&$ 3.5^{+15.2}_{ -2.0}$& 38 / 19&$6.8\times 10^{-3}$\\
132&$10.8^{+ 1.3}_{-1.3}$&$1.2^{+0.6}_{-0.6}$&$ 1.2^{+ 0.4}_{ -0.3}$& 39 / 30&0.12\\
\hline
{\bf NELGs:}&&&&&\\
&&&&&\\
100&$ 6.6^{+14.3}_{-4.3}$&$0.35^{+0.12}_{-0.12}$&$29^{+49}_{-20}$&  1 /  2&0.51\\
109&$12.2^{+ 1.7}_{-1.6}$&$0.72^{+0.17}_{-0.17}$&$ 5.9^{+ 1.4}_{ -1.1}$& 28 / 23&0.20\\
175&$ 3.5^{+ 1.5}_{-1.4}$&$4.0^{+0.4}_{-0.5}$&$ 2.5^{+ 2.8}_{ -1.3}$& 54 / 33&$1.2\times 10^{-2}$\\
225&$ 1.9^{+ 0.5}_{-1.9\dag}$&$0.00\dag^{+2.3}_{-0.0\dag}$&$ 0^{+1000\dag}_{-0\dag}$       & 18 /  3&$4.5\times 10^{-4}$\\
 \end{tabular}
\end{table*}

\begin{table*}
\caption{Absorbed power law fits with an additional soft emission line
component, modelled as a `mekal' thermal plasma.  $A$ is the normalisation of
the power law component in units of
$10^{-6}$~photons~cm$^{-2}$~s$^{-1}$~keV$^{-1}$. norm$_{mek}$ is the
normalisation of the thermal plasma given as $\int n_{e} n_{H} dV / $($4\pi
D_{l}^{2}$) in units of $10^{-20}$~cm$^{-5}$, where $n_{e}$ and $n_{H}$ are the
electron and hydrogen number densities respectively, $V$ is volume, and $D_{l}$
is the luminosity distance. Parameters and uncertainties marked with $\dag$ are
constrained by the limits of the fit range. P is the null hypothesis
probability corresponding to $\chi^{2}/\nu$.}
\label{tab:mekalfits}
 \begin{tabular}{lcccccl}
 &&&&&\\
\hline
 Source&$A$&$N_{H}$&$kT$&norm$_{mek}$&$\chi^{2}/\nu$&P\\
 &&($10^{22}$~cm$^{-2}$)&(keV)&&\\
\hline
{\bf BLAGN:} &&&&&&\\
&&&&&&\\
101            &$  2.9^{+ 0.6}_{ -0.6}$&$  0.70^{+  0.38}_{  -0.30}$&$  0.21^{+  0.04}_{ -0.03}$&$  1.7^{+  0.6}_{ -0.4}$& 19 / 18&0.39\\
132            &$ 11.7^{+ 1.2}_{ -1.2}$&$  0.94^{+  0.21}_{  -0.18}$&$  0.16^{+  0.08}_{ -0.06}$&$  2.1^{+ 12.3}_{ -1.4}$& 33 / 29&0.30\\
\hline
{\bf NELGs:} &&&&&&\\
&&&&&&\\
100            &$  6.5^{+23.0}_{ -4.4}$&$ 29^{+ 94}_{ -19}$&$ 4.3^{+  5.8}_{ -4.1}$&$  1.1^{+  0.5}_{ -0.5}$&  4 /  1&$5.4\times 10^{-2}$\\
109            &$ 11.7^{+ 2.4}_{ -1.6}$&$  6.7^{+  1.8}_{  -1.3}$&$ 10.0^{+  0.0\dag}_{ -5.2}$&$  2.7^{+  0.6}_{ -1.2}$& 28 / 22&0.18\\
175            &$  6.7^{+ 1.0}_{ -0.9}$&$  0.41^{+  0.22}_{  -0.16}$&$  0.19^{+  0.04}_{ -0.03}$&$  3.4^{+  1.3}_{ -0.8}$& 41 / 32&0.14\\
225            &$  0.9^{+ 7.5}_{ -0.8}$&$  0.02^{+ 11.50}_{  -0.02}$&$  0.77^{+  0.40}_{ -0.17}$&$  1.6^{+  1.1}_{ -0.8}$&  1 /  2&0.55\\
 \end{tabular}
\end{table*}

The alternative to an additional soft component, is that the absorbing 
material is ionised, in which case the strongest
absorption features will occur at higher energies than in a cold absorber, and
there may be no need for additional soft X-ray components in sources 101 and 132. Therefore, we
replaced the component of cold absorption with an ionised absorber (`absori' in
{\small XSPEC}), and refitted the spectra of sources 101 and 132, again assuming
$\Gamma=1.9$ and Galactic abundances. The results are given in Table
\ref{tab:absorifits} and show that this model fits both spectra well.

\begin{table*}
\caption{Power law fits with fixed $\Gamma = 1.9$ and an ionised absorber. $A$
is the power law normalisation in units of 
$10^{-6}$~photons~cm$^{-2}$~s$^{-1}$~keV$^{-1}$. $\xi$ is the ionisation
parameter in units of erg~cm~s$^{-1}$. Parameters and uncertainties marked with $\dag$ are
constrained by the limits of the fit range. 
P is the null hypothesis probability
corresponding to $\chi^{2}/\nu$.}
\label{tab:absorifits}
\begin{tabular}{lccccl}
 &&&&&\\
\hline
Source&$A$&$\xi$&$N_{H}$&$\chi^{2}/\nu$&P\\
&&&($10^{22}$~cm$^{-2}$)&&\\
\hline
{\bf BLAGN:}&&&&&\\
&&&&&\\
101&$3.3^{+0.6}_{-0.6}$&$140^{+180}_{-80}$&$2.6^{+2.4}_{-1.4}$&21 / 19&0.35\\
132&$12.0^{+1.2}_{-1.2}$&$3.9^{+5.7}_{-3.0}$&$1.4^{+0.4}_{-0.3}$&34 / 30&0.28\\
\hline
{\bf NELGs:}&&&&&\\
&&&&&\\
100&$2.1^{+1.7}_{-1.2}$&$60^{+120}_{-40}$&$7^{+9}_{-5}$&12 / 2&$3.0\times 10^{-3}$\\
109&$11.8^{+1.6}_{-1.5}$&$3.8^{+5.7}_{-3.0}$&$6.8^{+1.6}_{-1.3}$&59 / 23&$5.2\times 10^{-5}$\\
175&$7.3^{+1.0}_{-0.9}$&$60^{+80}_{-40}$&$1.3^{+0.9}_{-0.6}$&34 / 33&0.40\\
225&$1.9^{+0.6}_{-0.5}$&$0\dag^{+1000\dag}_{-0\dag}$&$0\dag^{+50}_{-0\dag}$&18 / 3&$4.5\times 10^{-4}$\\
&&&&&\\
\end{tabular}
\end{table*}

Absorption, either by cold or ionised gas, is unlikely to explain the poor fit
of the power law model to sources 52, 65 and 113, which have relatively soft
spectra (best fit $\Gamma > 2$). In all three cases, the data systematically
exceed the model at energies $\ge 2$~keV, indicating that the spectra are
curved relative to a power law, or that there is more than one broadband
component in the spectrum. We have explored two models for these spectra: a
power law with a reflection component, and a two power law model, which will
produce a curved continuum. The results of the reflection fits are given in
Table \ref{tab:reflec_fits}. In all 3 cases, the $\chi^{2}$ values indicate
acceptable fits when the reflection component is included, but the degree of
reflection required is far too large to be physically realistic.  The results
of the double power law fits are given in Table \ref{tab:double_powerfits}. The
goodness of fit is acceptable in all 3 cases.

\begin{table*}
\caption{Power law and cold reflection model fits to X-ray spectra of
BLAGN. $A$ is the power law normalisation in units of 
$10^{-6}$~photons~cm$^{-2}$~s$^{-1}$~keV$^{-1}$.
The reflection fraction $R$ is the size of the reflection
component relative to a plane reflector covering 2$\pi$ steradians.
P is the null hypothesis probability
corresponding to $\chi^{2}/\nu$.}
\label{tab:reflec_fits}
\begin{tabular}{lccccl}
 &&&&&\\
\hline
Source&$\Gamma$&$A$&$R$&$\chi^{2}/\nu$&P\\
\hline
&&&&&\\
52&$3.60^{+0.25}_{-0.22}$&$2.0^{+0.5}_{-0.5}$&$125^{+157}_{-75}$&41 / 29&$6.6 \times 10^{-2}$\\
65&$2.58^{+0.13}_{-0.11}$&$13.0^{+1.1}_{-1.2}$&$13.8^{+8.1}_{-5.3}$&93 / 97&0.59\\
113&$2.31^{+0.14}_{-0.14}$&$6.5^{+0.5}_{-0.6}$&$12.7^{+10.2}_{-6.5}$&63 / 55&0.21\\
\end{tabular}
\end{table*}

\begin{table*}
\caption{Fits with two power laws to X-ray spectra of BLAGN. $A_{1}$ and
$A_{2}$ are the normalisations in units of 
$10^{-6}$~photons~cm$^{-2}$~s$^{-1}$~keV$^{-1}$ for 
the power laws with photon indices $\Gamma_{1}$ and $\Gamma_{2}$ respectively.
P is the null hypothesis probability
corresponding to $\chi^{2}/\nu$.}
\label{tab:double_powerfits}
\begin{tabular}{lcccccl}
 &&&&&\\
\hline
Source&$\Gamma_{1}$&$A_{1}$&$\Gamma_{2}$&$A_{2}$&$\chi^{2}/\nu$&P\\
\hline
&&&&&&\\
52&$3.65^{+0.36}_{-0.26}$&$2.0^{+0.5}_{-0.7}$&$1.03^{+0.81}_{-0.89}$&$0.47^{+0.73}_{-0.36}$&42 / 28&$4.3 \times 10^{-2}$\\
65&$2.97^{+0.60}_{-0.35}$&$7.4^{+4.6}_{-4.5}$&$1.57^{+0.31}_{-0.48}$&$7.0^{+4.6}_{-4.3}$&96 / 96&0.48\\
113&$2.69^{+0.88}_{-0.40}$&$3.8^{+2.6}_{-2.9}$&$1.37^{+0.42}_{-0.79}$&$2.8^{+1.9}_{-2.3}$&61 / 54&0.24\\
\end{tabular}
\end{table*}

\subsection{NELGs}
\label{sec:nelgs}

The power law model was fitted to the NELG X-ray spectra and yielded acceptable
fits in 19/25 cases (Table \ref{tab:pofits}).  Of the 6 NELGs for which the
power law model is rejected at $>99\%$ confidence, 4 have best fit $\Gamma <
1$, suggesting that the poor fits are due to photoelectric absorption.  A
further 7 NELGs with acceptable power law spectral fits have best fit $\Gamma
< 1$. Such a hard spectral shape is unlikely to be the intrinsic continuum of
an AGN. Therefore we refitted all of the sources which are not well fitted by
the power law model, and all of the sources with best fit $\Gamma < 1$, with a
$\Gamma=1.9$ power law absorbed by a neutral column of material at the same
redshift as the object. The results are given in Table
\ref{tab:nhfits}. Acceptable fits are found for all of the sources except for
100, 109, 175, and 225. We refitted these four spectra using the same set of
models as for the BLAGN which could not be fitted with the cold absorption
model. The fit with an additional unabsorbed power law yields an acceptable
$\chi^{2}$ for sources 100, 109 and 175, but not for source 225 (Table
\ref{tab:popofits}), while the fit with a soft emission line component produces
acceptable $\chi^{2}$ for all 4 objects (Table \ref{tab:mekalfits}).  The
ionised absorption model is less successful: it produces a reasonable fit for
source 175 but not for the other three objects (Table \ref{tab:absorifits}).
Source 225 can only be adequately fitted when a soft thermal component is
included, and in this case there is no requirement that the power law component
is absorbed.

Of course it is possible that the NELGs which are well fitted by a $\Gamma> 1$ power law
 and no intrinsic absorption could still harbour absorption at a level that 
is too
small to be detected in our X-ray spectra. Therefore, for these 12 NELGs we
have determined upper limits to the intrinsic column densities by fitting
with a power law and intrinsic absorption, and increasing the column density
until $\Delta \chi^{2}=4$, corresponding to the 95\% upper limit. To be
conservative, we allowed $\Gamma$ to vary freely up to a maximum of
$\Gamma=2.5$. The results are given in Table \ref{tab:nelgaluppers};
the majority of the sources have upper limits to $\nh$ which are $\le 3\times
 10^{21}$~cm$^{-2}$. 

\subsection{Absorption line galaxies}
\label{sec:galaxies}

Of the 6 absorption line galaxies, only source 92 is poorly fitted with the power
law model (see Table \ref{tab:pofits}). This object, and source 98, require
very hard spectral indices ($\Gamma < 1$), so we refitted these two spectra with a
$\Gamma = 1.9$ power law and cold photoelectric absorption. In both cases this
provides an acceptable fit, as seen in Table \ref{tab:nhfits}.

For the four sources which were well fitted with a ($\Gamma > 1$) power law 
model, we have worked out 95\% upper limits to the amount of absorption 
which could be
present by adding intrinsic photoelectric absorption to the power law model,
and increasing the column density until $\Delta \chi^{2}=4$; the results are
given in Table \ref{tab:nelgaluppers}. 

\begin{table}
\caption{95\% upper limits to the intrinsic column densities of the NELGs and
galaxies which
are well fit with a $\Gamma>1$ power law model. 
}
\label{tab:nelgaluppers}
\begin{tabular}{lc}
 &\\
\hline
Source&$\nh$ 95\% upper limit\\
&($10^{22}$cm$^{-2}$)\\
\hline
{\bf NELGs:}&\\
&\\
1&0.17\\
20&0.026\\
56&0.077\\
94&3.6\\
103&0.30\\
115&0.12\\
118&0.68\\
131&0.50\\
153&1.68\\
186&0.23\\
193&0.26\\
200&0.62\\
\hline
{\bf galaxies:}&\\
&\\
37&0.096\\
55&0.22\\
117&3.1\\
184&0.070\\
\end{tabular}
\end{table}

\subsection{Galactic stars}
We began with a power law fit to the stars including full Galactic absorption,
exactly as for the extragalactic sources. The results are listed in Table
\ref{tab:pofits}, and the spectra are shown in Fig. \ref{fig:starspecs}. The
power law model is rejected in four of the five stars, which show a distinctive
emission peak between 0.5 and 1 keV, indicating optically-thin thermal
emission. However, the power law model is an acceptable fit to the X-ray
spectrum of source 179, which has by far the lowest quality spectrum.  We next
applied a single-temperature optically-thin thermal model (mekal). This model
also failed to produce acceptable fits for 4 out of 5 stars (see Table
\ref{tab:star1tempfits}): the stellar X-rays are distributed over a larger
energy range than the model spectra. Therefore we fitted a multi-temperature
mekal model that has a uniform distribution of emission measure 
with temperature up
to a maximum temperature $T$. The results of this model fit are listed in Table
\ref{tab:starmultitempfits} and shown in Fig. \ref{fig:starspecs}. This model
produced acceptable fits to the X-ray spectra of all the stars except source
140.  However, inspection of Fig. \ref{fig:starspecs} indicates that despite
the large $\chi^{2}$, the model reproduces the shape of the spectrum of source
140 quite well, but not in detail.  Source 140 has the highest signal/noise
X-ray spectrum by far within the \ukdeep\ field; a more complex model would be
required to reproduce the details, but this is beyond the scope of this paper.

\begin{table*}
\caption{Single-temperature optically-thin thermal plasma (mekal) fits to the
\xmm\ spectra of the Galactic stars. norm$_{mek}$ is the normalisation of
the thermal plasma given as $\int n_{e} n_{H} dV / $($4\pi D^{2}$) in units of
$10^{-20}$~cm$^{-5}$, where $n_{e}$ and $n_{H}$ are the electron and hydrogen
number densities respectively, $V$ is volume, and $D$ is the
distance. Parameters and uncertainties marked with $\dag$ are constrained by
the limits of the fit range.  
P is the null hypothesis probability
corresponding to $\chi^{2}/\nu$.}
\label{tab:star1tempfits}
\begin{center}
\begin{tabular}{lcccl}
 &&&&\\
\hline
Source&$kT$&$norm_{mek}$&$\chi^{2}/\nu$&P\\
&&(keV)&&\\
\hline
&&&&\\
49&$2.44^{+0.77}_{-0.49}$&$7.6^{+1.0}_{-1.1}$&51 / 19&$8.6\times 10^{-5}$\\
61&$0.26^{+0.01}_{-0.02}$&$3.0^{+0.3}_{-0.3}$&43 / 18&$8.3\times 10^{-4}$\\
120&$0.29^{+0.05}_{-0.03}$&$2.3^{+0.4}_{-0.4}$&37 / 11&$1.1\times 10^{-4}$\\
140&$0.68^{+0.01}_{-0.01}$&$192^{+3}_{-2}$&9645 / 165&$< 10^{-30}$\\
179&$2.9^{+17.1\dag}_{-1.6}$&$1.6^{+1.3}_{-0.8}$&3 / 4&0.56\\
\end{tabular}
\end{center}
\end{table*}

\begin{table*}
\caption{Multi-temperature optically-thin thermal plasma (cemekl) fits to the
\xmm\ spectra of the Galactic stars, with emission measure distributed
uniformly with temperature for $kT < kT_{max}$. norm$_{mek}$ is the
normalisation of the thermal plasma given as $\int n_{e} n_{H} dV / $($4\pi
D^{2}$) in units of $10^{-20}$~cm$^{-5}$, where $n_{e}$ and $n_{H}$ are the
electron and hydrogen number densities respectively, $V$ is volume, and $D$ is
the distance. Parameters and uncertainties marked with $\dag$ are constrained
by the limits of the fit range.  
P is the null hypothesis probability
corresponding to $\chi^{2}/\nu$.}
\label{tab:starmultitempfits}
\begin{center}
\begin{tabular}{lcccl}
 &&&&\\
\hline
Source&$kT_{max}$&$norm_{mek}$&$\chi^{2}/\nu$&P\\
&&(keV)&&\\
\hline
&&&&\\
49&$20.0\dag^{+0\dag}_{-8.4}$&$2.3^{+0.3}_{-0.2}$&28 / 19&$9.3\times 10^{-2}$\\
61&$0.45^{+0.09}_{-0.06}$&$4.9^{+0.8}_{-0.8}$&20 / 19&0.37\\
120&$7^{+13\dag}_{-6}$&$2.2^{+0.5}_{-0.5}$&8 / 11&0.67\\
140&$4.72^{+0.30}_{-0.25}$&$236^{+3}_{-4}$&688 / 165&$< 10^{-30}$\\
179&$20\dag^{+0\dag}_{-19}$&$0.38^{+0.20}_{-0.13}$&6 / 4&0.24\\
\end{tabular}
\end{center}
\end{table*}

\section{Discussion}

\begin{figure}
\begin{center}
\leavevmode
\psfig{figure=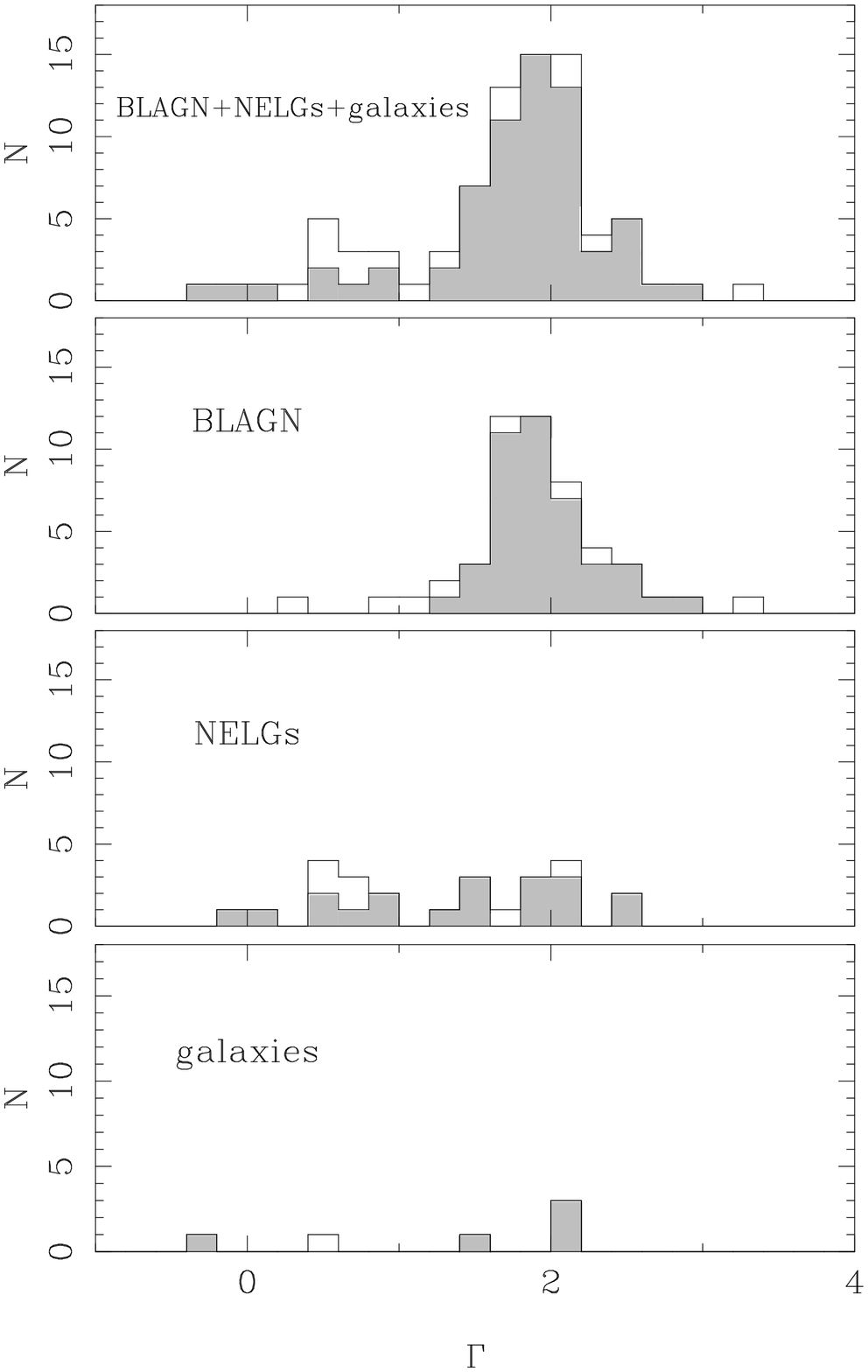,width=80truemm}
\caption{Best fit power law indices for the different extragalactic point
sources. The filled region corresponds to those sources which are 
acceptably fitted
with a single power law.}
\label{fig:gammahist}
\end{center}
\end{figure}

\subsection{Distinguishing Galactic stars from extragalactic objects}
The overall X-ray spectral shapes of the Galactic stars could be reproduced 
by an optically-thin thermal model with a distribution of temperatures. This is
broadly consistent with observations of nearby stars with active coronae
\citep[e.g. ][]{guedel01}. 
The characteristic shape of their X-ray spectra, 
and the statistical rejection of the power law model,
mean that these stars can be distinguished from the majority of the BLAGN by
their EPIC X-ray spectra alone.  This has considerable potential for the
identification of Galactic stars in \xmm\ surveys, independent of optical
spectroscopy. In the \ukdeep\ field, it provides confirmation that sources 49,
61, 120 and 140, have been correctly identified as Galactic stars. For source
141, the signal to noise of the X-ray spectrum is not sufficient to confirm or
reject its identification as a Galactic star. 

\begin{figure}
\begin{center}
\leavevmode
\psfig{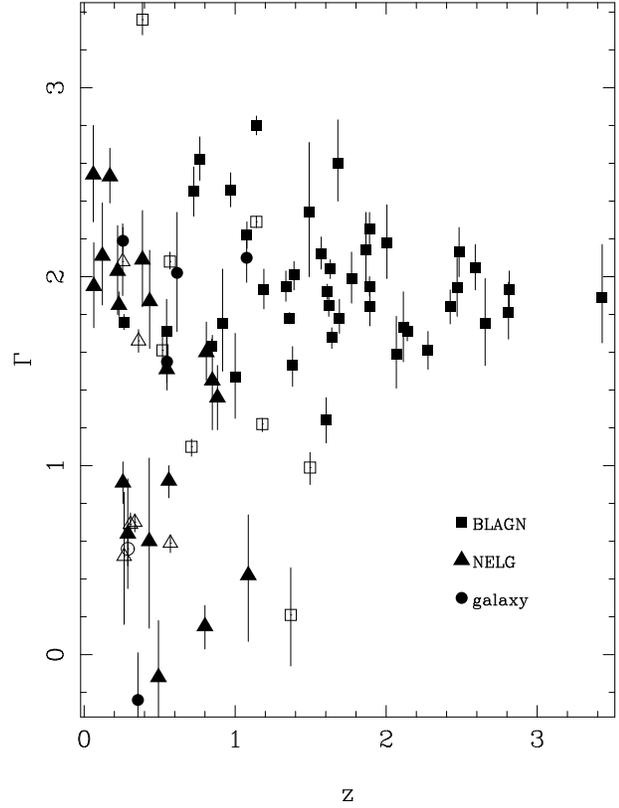}
\caption{Best fit power law slopes from single power law model fits, shown 
for the different source types as a function of
redshift. Poor fits are shown with open symbols.}
\label{fig:zdiffpofits}
\end{center}
\end{figure}

\subsection{Spectral slopes of the extragalactic sources}

The distribution of best fit power law slopes is shown in
Fig. \ref{fig:gammahist}. The distribution for all sources appears to be
bimodal, with the larger peak at $\Gamma \sim 1.9$ and the smaller peak at
$\Gamma \sim 0.6$. However, when the distribution is broken into BLAGN, NELGs
and galaxies it can be seen that the peak at $\Gamma \sim 0.6$ comes almost
entirely from the NELGs, of which a similar number have $\Gamma \sim 0.6$ as
have $\Gamma \sim 1.9$.  The BLAGN power law slopes show no indication of
bimodality. The power law slopes are shown as a function of redshift in
Fig. \ref{fig:zdiffpofits}. The slopes of the BLAGN show no trend with
redshift, and are distributed relatively symmetrically around 
$\Gamma \sim 1.9$ at all redshifts. The distribution of 
NELG and galaxy spectral slopes also shows no obvious trend with redshift. 

The most outlying BLAGN photon indices (at $\Gamma < 1$ and $\Gamma > 3$) come
from spectra which are not well fitted with a power law model. In contrast, the
majority of the NELG spectra with $\Gamma < 1$ are acceptably fitted 
with a power
law, perhaps because they have on average fewer counts in their X-ray spectra
than the BLAGN. Each of the 7 hard spectrum NELGs which have $\Gamma<1$ in the
power law fit can also be fitted with an absorbed ($\Gamma=1.9$) power law.
Although the individual spectra do not have good enough statistics to
discriminate between the two models, we can compare the goodness of fit for the
two models using the total $\chi^{2}/\nu$ for the 7 objects. For the power law
model, the total $\chi^{2}/\nu = 75/43$, corresponding to a null hypothesis
probability of $1.8\times 10^{-3}$, whereas the total $\chi^{2}/\nu = 66/43$
for the absorbed power law model, corresponding to a null hypothesis
probability of $1.4\times 10^{-2}$. Thus overall, the model with photoelectric
absorption produces a better fit to the spectra of these 7 hard-spectrum NELGs
than the unabsorbed power law model.

The sample of galaxies is much smaller than the sample of NELGs, and there are
only two objects with best fit $\Gamma<1$, one of which is acceptably fitted with
a power law model, and one of which is not. Both of these objects are fitted
well with an absorbed $\Gamma=1.9$ power law. Overall, these results suggest
that photoelectric absorption is responsible for all of the spectra which
appear to have $\Gamma<1$ in power law model fits.

\begin{figure}
\begin{center}
\leavevmode
\psfig{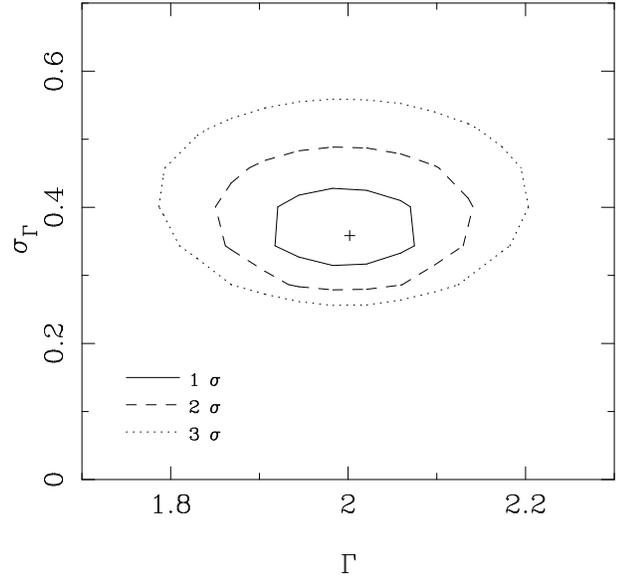}
\caption{Maximum likelihood confidence intervals for the mean spectral index
$\langle \Gamma \rangle$ and standard deviation $\sigma_{\Gamma}$ of BLAGN
power law spectral indices assuming a gaussian distribution. The three
contours correspond to 1, 2 and 3 sigma for two interesting parameters
(i.e. $\Delta S =$~2.3, 6.2, and 7.8 where $S=-2 \log_{e} L$ and $L$ is the
likelihood function).}
\label{fig:maccacaro}
\end{center}
\end{figure}

Only a small proportion (5/50) of the BLAGN show significant evidence for
absorption, hence for the BLAGN we can obtain a good estimate of the intrinsic
distribution of spectral slopes. For this we assume a Gaussian distribution of
spectral slopes and use the maximum likelihood method of \citet{maccacaro88}
which takes into account the statistical errors on the individual spectral
fits. We exclude the 5 sources which show evidence for photoelectric absorption
because it is difficult to determine the intrinsic spectral shape for
these sources. Three of the BLAGN appear to have spectra which are more complex
than a power law, but we include these objects with their best fit single power
law indices when calculating the distribution.  The likelihood contours for the
mean of the distributon $\langle \Gamma \rangle$ and the intrinsic dispersion
$\sigma_{\Gamma}$ are shown in Fig. \ref{fig:maccacaro}. The one-dimensional
confidence intervals are $\langle \Gamma \rangle = 2.00\pm0.11$,
$\sigma_{\Gamma} = 0.36^{+0.10}_{-0.07}$.

Our best fit $\langle \Gamma \rangle$ is entirely consistent with that found by
\citet{mateos05b} for a comparable sample of BLAGN
in the Lockman Hole, and with that found by \citet{mateos05a} for a larger, but brighter sample of BLAGN studied with
\xmm. Indeed, $\langle \Gamma \rangle = 2.00\pm0.11$ is also consistent with
the value of $\langle \Gamma \rangle = 2.05\pm0.05$ determined by
\citet{mittaz99} for AGN in the soft X-ray \ros\ International X-ray Optical
Survey (RIXOS). However, the value of $\sigma_{\Gamma}$ found in the \ukdeep\
field is somewhat larger than that found by \citet{mateos05a} and 
\citet{mateos05b}, who obtained
$\sigma_{\Gamma} = 0.21^{+0.05}_{-0.04}$ and $\sigma_{\Gamma} = 0.20\pm 0.04$
respectively (their confidence intervals are
90\%), and it is smaller than the $\sigma_{\Gamma}=0.55\pm0.05$ obtained 
by \citet{mittaz99} for
RIXOS AGN. It is likely that the large
dispersion found with \ros\ is due to the 
fact that photoelectric absorption and
AGN soft excess emission were both 
difficult to distinguish from a power law with
the crude \ros\ PSPC spectra. Furthermore, photoelectric absorption and AGN
soft excess emission both have a larger effect in the 0.1--2 keV band of the
PSPC than in the 0.2--12 keV band of EPIC.


\subsection{Spectra which are more complex than a power law with 
cold photoelectric
absorption}
\label{sec:complexity}

There were 5 BLAGN and 4 NELGs which show spectra which are not well fitted
with a power law and cold absorber model. Here we discuss these objects in more
detail, starting with the BLAGN.

It is likely that the poor fit of the power law model to sources 65 and 
113 owes more to the fact that they are among the best 
BLAGN spectra in terms of signal to noise ratio, than to a particularly large
degree of curvature relative to the other BLAGN
(Fig. \ref{fig:blagnspecs}). However, source 52 
appears to show a more substantial
departure from a single power law spectrum, showing a very soft spectrum 
below 2~keV,
but a much harder spectrum at higher energy. This source showed a very soft
spectrum in the original \ros\ survey \citep{mchardy98}, and is a luminous
example of a narrow line Seyfert 1. 
It was studied in its own right by \citet{dewangan01}. The spectrum of
source 52 in Fig. \ref{fig:blagnspecs}, shows significant similarity to the
spectrum of the nearby narrow-line Seyfert 1 galaxy 1H~0707-495.  In
particular, 1H~0707-495 also has a very steep spectrum ($\Gamma=3.8$) when fitted
with a single power law over the full \xmm\ energy range, but has a much
flatter spectrum above 2 keV \citep{boller02}. The most remarkable feature of
the spectrum of 1H~0707-495 is a large edge at 7-7.6 keV and a significant
deficit of counts at energies higher than this \citep{boller02,gallo04}. For
source 52 at $z=0.386$, a similar edge would be redshifted to 5-5.5 keV. Source
52 is not detected beyond this energy, so it could harbour a similar edge
feature to 1H~0707-495. However, the spectrum is not good enough for us to
determine with confidence whether such an edge is present at this energy.

Of the 5 BLAGN which show absorption, 3 are well fitted with a power law and
cold photoelectric absorption, while sources 101 and 132 require more
complicated models. Acceptable fits are found for these 2 sources 
when an additional soft emission
component is included, or when the neutral absorber is replaced by an ionised
one. In the model fits including an additional soft line emitting or unabsorbed
power law component, source 132 has a soft component which emits between 5\%
and 21\% as much power in the 0.2-1 keV band (observer frame) as the primary
power law component. For source 101 the additional soft component is much more
important, emitting 35--120\% as much radiation in the 0.2-1 keV band as the
primary power law. In both cases the additional soft component emits
$>10^{42}$~erg~s$^{-1}$ in the 0.2-1~keV band: such a high X-ray luminosity
exceeds even the most X-ray luminous starburst galaxy known \citep{moran99}.
It is also unlikely that the soft emission comes from a group or cluster of
galaxies in which the source is embedded because at such luminosities the X-ray
emitting gas would be expected to have $kT \sim 1$~keV \citep{helsdon00},
whereas the best fit temperatures are much lower than this in the mekal
fits. Furthermore, sources 101 and 132 are consistent with point sources in the
X-ray images, and are isolated, separated by more than an arcminute from their
nearest \xmm\ or \chandra\ neighbours. Therefore if these sources do have an
additional soft component, it must be scattered or reprocessed AGN
emission. While this is plausible for source 132, in which the soft component
is $\sim 10\%$ as luminous as the primary AGN emission, for source 101 a
scattered component would have a luminosity of more than a third of the primary
component, which is untenable.  The alternative, an ionised absorber, provides
a good fit to both spectra, and is the only realistic hypothesis for source
101.

The situation is quite different for the four absorbed NELGs which are not well
fitted with a power law and neutral photoelectric absorption. For three of these,
sources 100, 109, and 225, the ionised absorber model does not fit the data,
and an additional soft component is therefore neccessary.  In the case of
source 109, the additional soft component is likely to be due to contamination
of the X-ray spectrum from the bright BLAGN, source 113, which is only 20$''$
distant. Indeed, in the fit to source 109 with a mekal component, $kT$ pegs at
10~keV, which is the maximum value that we allowed in the fit. At such a high
temperature the mekal component is contributing more continuum than line
emission, consistent with the hypothesis that the excess flux in the lowest
energy channels of source 109 is actually continuum emission from source 113.

For source 225 the best fit mekal component has $kT=0.8$~keV, and a 0.2-1~keV
luminosity of $10^{42}$~ergs~s$^{-1}$, larger than that of the power law
component in the same band. This luminosity is too large for either a starburst
or scattered AGN emission to be plausible. However, source 225 is one of the
seven extended X-ray sources identified with `{\small EMLDETECT}'
\citep{loaring05}, with a FWHM extent of $\sim 17$ arcseconds, so a 
significant fraction of the X-ray emission is likely to
originate in a group or cluster of galaxies. The temperature and luminosity of
source 225 are consistent with the $L_{X}-T_{X}$ relationship observed in
galaxy groups \citep{helsdon00}, further supporting this explanation.  

In source 100 the soft component has a relatively low luminosity
($10^{41}$~ergs~s$^{-1}$ in the 0.2--1 keV band), equivalent to $\sim 5\%$ of
the primary power law emission in the same band. Therefore scattered emission
from the AGN and hot gas from a starburst are both tenable explanations for the
soft component in this source.

In the fit to source 175 including a mekal component, we obtain a very
reasonable $kT=0.2$~keV, but a soft component 0.2--1~keV luminosity of $2\times
10^{43}$~ergs~s$^{-1}$. This is equivalent to $30-50\%$ of the luminosity of
the power law component in the same band, hence neither a starburst
contribution, nor scattered AGN emission are feasible explanations. It is
possible that a nearby X-ray source contributes to the soft emission: {\em
Chandra} source number 159 from \citet{mchardy03} is only $11''$ distant, but
is a factor of 10 fainter than source 175, and a visual inspection of the \xmm\
0.2--0.5~keV image indicates that this source does not contribute
significantly. Therefore it appears that an ionised absorber is a better
hypothesis for the X-ray spectral shape of source 175 than an additional soft
component.

\begin{figure}
\begin{center}
\leavevmode
\psfig{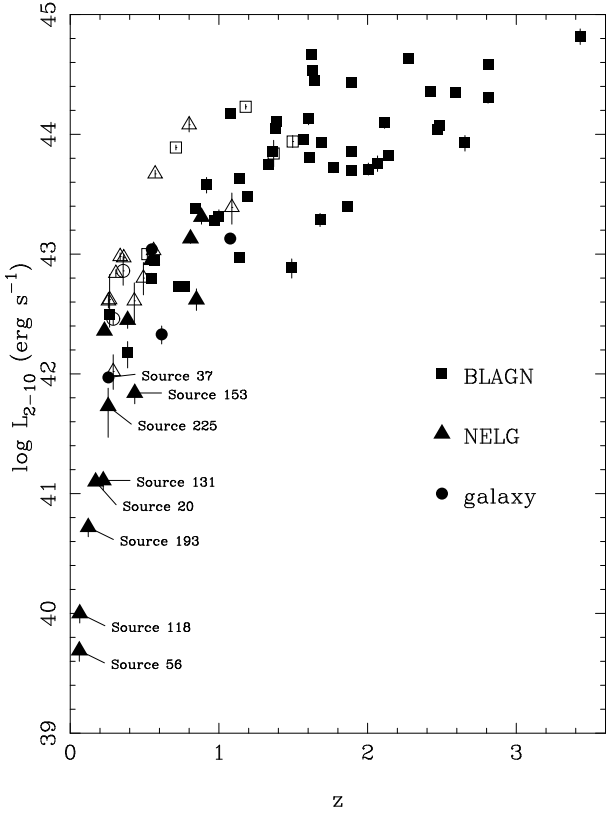}
\caption{2--10 keV intrinsic rest-frame luminosities as a function of 
redshift. The sources which
show X-ray absorption in their X-ray spectra are shown with open symbols, while
sources without significant X-ray absorption are shown with closed symbols.}
\label{fig:zbest21lumin}
\end{center}
\end{figure}

\subsection{The luminosity distribution as a function of optical class}

In Fig. \ref{fig:zbest21lumin} we show the 2--10 keV luminosities ($L_{2-10}$) 
as a function
of redshift for the extragalactic sources. The luminosities are for the primary
power law component only, are corrected for intrinsic and Galactic absorption,
and are derived from the most realistic model fit to each source, as discussed
in the text.
At $z>1$, optical spectroscopic identification of NELGs and galaxies is
extremely difficult compared to BLAGN, so it is not suprising that the vast
majority of optically identified sources at $z>1$ are BLAGN. Therefore it is
useful to restrict our attention to $z<1$, where we are able to identify the
full range of optical counterparts, in order to examine the luminosity and
absorption characteristics of the NELGs and galaxies with respect to the BLAGN.
For $z<1$, we see in Fig. \ref{fig:zbest21lumin} that the majority (13/15) of
the X-ray absorbed sources are NELGs or galaxies, as expected in AGN
unification schemes.  However, the majority (15/29) of the NELGs and 
galaxies in this redshift range (i.e. $z<1$) {\em do
not} show significant X-ray absorption. The NELGs and galaxies span the full
$z<1$ X-ray luminosity range, but the X-ray absorbed sources are only found at
$L_{2-10}> 10^{42}$~ergs~s$^{-1}$. Indeed, the X-ray absorbed sources have a
higher mean 2--10~keV luminosity ($\langle L_{2-10}\rangle = 2\times
10^{43}$~ergs~s$^{-1}$) than the unabsorbed sources, which have $\langle
L_{2-10}\rangle = 8\times 10^{42}$~ergs~s$^{-1}$.  
However the mean luminosity
of the NELGs and galaxies ($\langle L_{2-10}\rangle = 10^{43}$~ergs~s$^{-1}$)
is smaller than the mean luminosity of the BLAGN ($\langle L_{2-10}\rangle =
2\times 10^{43}$~ergs~s$^{-1}$).

\subsection{The origin of the X-ray emission in the low luminosity
extragalactic sources}
\label{sec:lowlum}

For sources with $L_{2-10}>10^{42}$~erg~s$^{-1}$, the presence of an AGN is
unambiguous. However, the lower luminosity X-ray sources could plausibly be
powered by star formation, in which case the X-rays will be a composite of
emission from massive X-ray binaries and optically thin thermal emission from
hot interstellar gas within the galaxy. Therefore to determine whether the
X-ray emission from these objects is powered by AGN, we must examine them in
more detail on a case by case basis.

The two lowest luminosity sources, 56 and 118, have $L_{2-10}$ of $5\times
10^{39}$ and $1\times 10^{40}$~ergs~s$^{-1}$ respectively. They are both 
detected as
point sources with \xmm\ but are not detected with our point source searching
method in the \chandra\ images \citep{mchardy03,loaring05},
suggesting that the X-ray emission is signficantly extended with respect to the
\chandra\ PSF (i.e. on a scale comparable to the optical galaxy). They are both
bright radio sources ($>0.5$~mJy at 1.4\,GHz), extended on a similar scale to
the optical galaxy \citep{seymour04} and so almost certainly powered by star
formation.  The starburst nature of source 56 is discussed at length in
\citet{gunn01}.  In Fig. \ref{fig:nelgspecs} we see that although the spectra
are consistent with a power law shape, both have a low significance peak at
$\sim 0.7$~keV, consistent with thermal emission from a starburst.  Thus the
X-ray emission from both source 56 and source 118 could quite plausibly be
powered by star formation.

In contrast, 
the sources with $L_{2-10} > 2\times 10^{40}$~ergs~s$^{-1}$
are much more likely to be AGN powered. Sources 20, 131 and 193 (all NELGs) 
show no
evidence for thermal gas components in their spectra, which are consistent with
power laws, suggesting that they are AGN.  Both sources 131 and 193 are
detected as point sources in the \chandra\ images \citep{mchardy03}, consistent
with this hypothesis. Source 20 was not detected with \chandra, but in 
this
case variability, rather than spatial extent, is responsible. Source 20 was
more than a factor 2 brighter during \xmm\ revolution 0276 than it was 11 days
later in revolution 282, and hence the emission must come from an AGN rather
than a multitude of sources within the galaxy.

At higher luminosities, the hypothesis that star formation powers the
X-ray emission becomes even less likely. Source 225 has a thermal component in
its X--ray spectrum, but as discussed in Section \ref{sec:complexity} the X-ray
emission is extended on a scale larger than the optical galaxy in this source,
so the thermal emission is due to a group or cluster of galaxies.  Source 153
is not detected at 1.4\,GHz \citep{seymour04}, 
so is unlikely to be a vigourously star forming
galaxy, while source 37 is almost certainly a BL Lac object \citep[][Moss
et~al. in prep]{gunn03}.

From the case by case examination of the 8 sources with $L_{2-10} <
10^{42}$~ergs~s$^{-1}$, only the two lowest
luminosity objects, 56 and 118, which have $L_{2-10} \le
10^{40}$~ergs~s$^{-1}$, appear to have X-ray emission dominated by star
formation. In all the other cases the 2--10~keV 
X-ray emission appears to be AGN
dominated. Apart from source 225, the AGN is also dominant in the 0.5--2 keV
band in these objects.

\subsection{The nature of the NELGs and galaxies without X-ray absorption}

Excepting objects which are powered by star formation, the standard paradigm
for X-ray selected NELGs and absorption line galaxies is that they are obscured
AGN \citep[e.g. ][]{hasinger01,mainieri02,barcons02}.  However in the \ukdeep\
field, a significant number of the X-ray sources which are optically classified
as NELGs and galaxies contain AGN (see Section \ref{sec:lowlum}) which do {\em
not} show evidence for X-ray absorption. While for three of these 
objects (sources
94, 153 and 117), the current data do not allow us to rule out 
 an absorber with a significant column density ($\nh >
10^{22}$~cm$^{-2}$), for more than than half (8/14) of these objects 
the data imply $\nh \le 3\times 10^{21}$~cm$^{-2}$ 
(Table \ref{tab:nelgaluppers}, excluding objects 56 and 118 which are likely 
powered by star formation). For a Galactic gas to dust ratio and reddening
curve \citep{bohlin78}, this would correspond to at most a 
factor of 3 attenuation to the broad H$\alpha$ line. For some of the lower 
redshift objects the $\nh$ limits are so restrictive that optical reddening of
the AGN should be negligible, e.g. source 20 has $\nh < 2.6\times
10^{20}$~cm$^{-2}$. \citet{silverman05} reported a number
of objects in the \chandra\ Multiwavelength Project which appear to 
have similar
properties and argued that this can be attributed to dilution of the AGN
emission by the host
galaxy light \citep[see also][]{pageetal03,severgnini03}.

\begin{figure}
\begin{center}
\leavevmode
\psfig{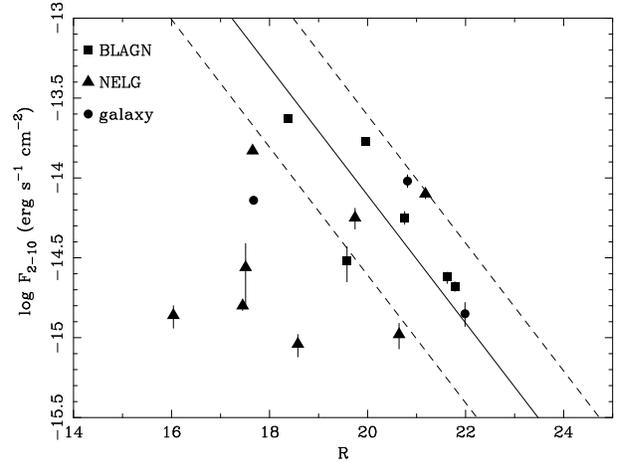}
\caption{2-10 keV flux against R
magnitude, for unabsorbed X-ray sources with $z<0.8$. The solid line
corresponds to the unweighted mean X-ray/optical ratio of the BLAGN shown, and
the dashed lines delimit $\pm$0.5 dex around this ratio.}
\label{fig:flux21vsR}
\end{center}
\end{figure}

The dilution 
hypothesis can be investigated by comparing the optical to X-ray flux
ratios as a function of optical classification. The X-ray flux is completely
dominated by the AGN component, while the optical flux comes largely from the
AGN in sources classified as BLAGN but is dominated by the host galaxy in
sources classified as NELGs and galaxies (Loaring et~al. in prep).
Fig. \ref{fig:flux21vsR} shows the 2--10 keV flux of unabsorbed galaxies and
NELGs with $z<0.8$,
against the R-band optical magnitude
from \citet{mchardy03}.  Sources 56 and 118 are not shown, because their X-ray
emission is likely to be powered by stellar processes. Fig. \ref{fig:flux21vsR}
is restricted to sources with $z< 0.8$ so that the R band corresponds to rest
frame optical emission longward of the Balmer break. The solid line in
Fig. \ref{fig:flux21vsR} corresponds to the mean optical/X-ray ratio of the 6
BLAGN that are shown.  The dashed lines correspond to $\pm0.5$~dex about this
ratio, which is approximately the standard deviation of optical/X-ray ratios
found for X-ray selected BLAGN \citep{ciliegi95}. The NELGs are distributed
over a broader region of Fig. \ref{fig:flux21vsR} than the BLAGN, and more than
half of the NELGs lie to the left of the region enclosed by the dashed lines,
i.e. they have a larger optical/X-ray flux ratio than most of the BLAGN. Four
of the NELGs are more than 3 magnitudes to the left of the solid line in
Fig. \ref{fig:flux21vsR}. Therefore if they host active nuclei with
optical/X-ray flux ratios typical for the BLAGN in Fig. \ref{fig:flux21vsR},
the nuclear component will be outshone by more than a factor of 15 in the
optical by the surrounding galaxy. Two of the NELGs and one galaxy lie 1.5--2.5
magnitudes to the left of the solid line, suggesting that the active nucleus
will be outshone by the host galaxy by factors of 3--10; dilution of the AGN
component by starlight could reasonably explain the absence of broad emission
lines in the optical spectra of these objects. The four remaining NELGs and
galaxies, which lie within the region bounded by the dashed lines, have
optical/X-ray ratios similar to those of the BLAGN. In these objects the
stellar component produces a similar flux of optical radiation as we would
expect to observe from the nucleus, given the X-ray flux. For dilution of the
nuclear spectrum by galaxy starlight to explain the optical characteristics of
these sources, their nuclear components must have optical/X-ray flux ratios
which are smaller than the average for BLAGN. This however, is quite plausible
if AGN with $L_{2-10} < 10^{43}$~erg~s$^{-1}$ 
have a similar intrinsic scatter of
optical/X-ray flux ratios to the scatter found in more luminous BLAGN
\citep[$\sim 0.5$~dex,][]{ciliegi95}. The lack of NELGs and galaxies to the
right of the dashed-line region of Fig. \ref{fig:flux21vsR} is consistent with
this picture: starlight is unlikely to render the broad emission lines
unidentifiable when the host galaxy has a low luminosity compared to the
nucleus, unless the optical/X-ray flux ratio of the nucleus is unusually low.

Fig. \ref{fig:flux21vsR} therefore implies that in the NELGs and galaxies which
do not show significant X-ray absorption, the lack of broad optical emission
lines could be due to the low contrast of the emission lines against the much
stronger starlight component. However, it should be noted that there are
alternative hypotheses that are not presently ruled out. These objects could
posess absorbing media with a high dust to gas ratio, and/or in which the gas
is ionised and so its X-ray photoelectric opacity is reduced \citep{pappa01}.
Discrimination between these hypotheses will require high spatial resolution,
high signal to noise optical spectra \citep{severgnini03}.  Alternatively, some
of these objects could be BL Lac objects; this appears to be the case for at
least source 37 \citep[][Moss
et~al. in prep]{gunn03}.

We calculated the absolute magnitudes $M_{R}$ for the 11 NELGs and galaxies
shown in Fig.~\ref{fig:flux21vsR}. 
The results are given in Table \ref{tab:absmags}, and
were calculated using the elliptical galaxy K-corrections from
\citet{coleman80}. These objects lie in the bright part of the galaxy
luminosity function, having absolute magnitudes of around $M^{*}_{R}$ or
brighter, where $M^{*}_{R}$ \citep[$\sim 21.8$ 
for $H_{0}=70$~km~s$^{-1}$~Mpc$^{-1}$,
][]{trentham05} defines the knee in the luminosity function. On the other hand,
they all host AGN which have $L_{2-10} < 2\times 10^{43}$~erg~s$^{-1}$),
lower than the knee in the X-ray
luminosity function \citep[e.g. ][]{page97,barger05,hasinger05}. This
combination of weak AGN and luminous host galaxies, given the
$M_{\bullet}-\sigma$ relation \citep{ferrarese01}, suggests that these objects
host massive black holes \citep[$\ge 10^{8}$\,M$_{\odot}$,][]{page01b} with low
Eddington ratios ($<1\%$). Such objects are not low mass objects formed at late
times, the result of AGN `cosmic downsizing' as argued by \citet{barger05}, but
are instead the relatively inactive remnants of once-luminous AGN.

\begin{table}
\caption{Absolute magnitudes $M_{R}$ for the $z<0.8$ galaxies and NELGs that
contain an AGN, and which do not show significant X-ray absorption. }
\label{tab:absmags}
\begin{center}
\begin{tabular}{lc}
 &\\
\hline
Source  &   $M_{R}$ \\
\hline
&\\
1    &  -22.306 \\ 
20    &  -22.303 \\ 
37    &  -23.175 \\ 
103    &  -22.169 \\ 
115    &  -22.892 \\ 
117    &  -21.896 \\ 
131    &  -21.872 \\ 
153    &  -21.814 \\ 
184    &  -22.550 \\ 
193    &  -22.834 \\ 
225    &  -23.317 \\ 
\end{tabular}
\end{center}
\end{table}

\section{Conclusions}

We have presented the \xmm\ EPIC X-ray spectra of 86 X-ray sources in the
\ukdeep\ field which have optical identifications and $> 70$ X-ray counts.
More than half of the sources are BLAGN, and the majority of these (42/50) 
have spectra which are consistent with a power law shape. The photon indices
are distributed about a mean $\langle \Gamma \rangle = 2.0\pm0.1$ with an 
intrinsic
dispersion of $\sigma_{\Gamma} = 0.4\pm0.1$. Of the 8 BLAGN which have X-ray
spectra inconsistent with a power law shape, 3 have curvature with excess 
emission at the high and low energy ends relative to a
power law, and 5 show evidence for absorption at soft X-ray energies. 
The addition of a cold photoelectric absorber yields an acceptable fit for 3 
of these objects, one requires an ionised absorber, and the remaining object
can be fitted with either an ionised absorber or with an additional component 
of
scattered or reprocessed soft X-ray emission.

Of the 25 NELGs, 13 have absorbed X-ray spectra, 9 of which can be fitted with
a cold photoelectric absorber, and 4 of which require a more complex model with
either an ionised absorber or an additional component of soft X-ray emission. 
Of the 6 absorption line galaxies in the sample, 2 show evidence for X-ray
absorption, and both are consistent with a simple cold photoelectric absorber.
Of the 12 NELGs which do not show evidence for X-ray absorption, 2 have X-ray
properties which suggest that their X-ray emission is powered by star 
formation. The other 14 NELGs and galaxies which are not X-ray absorbed 
have X-ray properties implying the presence of an
AGN. At least at $z<0.8$, where our R-band imaging probes the rest-frame
optical starlight, these objects are luminous galaxies 
($M_{R}>M_{R}^{*}$), containing
relatively low X-ray luminosity AGN ($L_{2-10}<2\times 10^{43}$~erg~s$^{-1}$), 
and 
the lack of AGN
signatures in the optical could be due to the dilution of the AGN radiation by
the bright host galaxy.  These objects are likely to be massive black holes
($>10^{8}$\,M$_{\odot}$) accreting 
with low Eddington ratios ($<0.01$), rather than vigorous, low mass black 
holes.

Finally, the sample contains 5 Galactic stars which have X-ray spectra
consistent with optically thin thermal emission with a broad distribution of
temperature, consistent with active stellar coronal emission. In 4 of the 5
stars, the X-ray spectra alone can distinguish them from AGN at similar X-ray
flux levels.

\section{Acknowledgments}
Based on observations obtained with XMM-Newton, an ESA science mission with
instruments and contributions directly funded by ESA Member States and NASA.
The William Herschel Telescope is operated on the island of La Palma by the 
Isaac Newton
Group in the Spanish Observatorio del Roque de los Muchachos of the Instituto
de Astrofisica de Canarias. The W.M. Keck Observatory is operated as a
scientific partnership among the California Institute of Technology, the
University of California and the National Aeronautics and Space Administration
and was made possible by the generous financial support of the W.M. Keck 
Foundation. We thank Sergey Kuznetsov for helpful comments.


\begin{thebibliography}{}

\bibitem[Alexander et~al.(2003)]{alexander03}
Alexander D.M., et~al., 
2003, AJ, 126, 539

\bibitem[Anders \& Grevesse(1989)]{anders89} Anders E. \& 
Grevesse N., 1989, Geochimica et 
Cosmochimica Acta, 53, 197

\bibitem[Antonucci(1993)]{antonucci93} Antonucci R., 1993, 
Annu. Rev. A\&A, 31, 473

\bibitem[Barcons et~al.(2002)]{barcons02}
Barcons X., et~al.,
2002, A\&A, 382, 522

\bibitem[Barcons et~al.(2003)]{barcons03}
Barcons X., Carrera F.J., Ceballos M.T., 
2003, MNRAS, 339, 757

\bibitem[Barger et~al.(2003)]{barger03}
Barger A.J., et~al.,
2003, AJ, 126, 632

\bibitem[Barger et~al.(2005)]{barger05}
Barger A.J., Cowie L.L., Mushotzky R.F., Yang Y.,
Wang W.-H., Steffen A.T., Capak P.,
2005, AJ, 129, 578

\bibitem[Bohlin et~al.(1978)]{bohlin78}
Bohlin R.C., Savage B.D., Drake J.F.,
1978, ApJ, 224, 132

\bibitem[Boller et~al.(2002)]{boller02}
Boller Th., et~al., 
2002, MNRAS, 329, L1

\bibitem[Caccianiga et~al.(2004)]{caccianiga04}
Caccianiga A., et~al., 2004, A\&A, 416, 901

\bibitem[Ciliegi et~al.(1995)]{ciliegi95}
Ciliegi P., Elvis M., Wilkes B.J., Boyle B.J., McMahon R.G., Maccacaro T.,
1995, MNRAS, 277, 1463

\bibitem[Coleman, Wu \& Weedman(1980)]{coleman80}
Coleman G.D., Wu C.-C., Weedman D.W.,
1980, ApJS, 43, 393

\bibitem[Comastri et~al.(1995)]{comastri95}
Comastri A., Setti G., Zamorani G., Hasinger G.,
1995, A\&A, 296, 1

\bibitem[Crawford et~al.(2002)]{crawford02}
Crawford C.S., Gandhi P., Fabian A.C., Wilman R.J.,
Johnstone R.M., Barger A.J., Cowie L.L.,
2002, MNRAS, 333, 809

\bibitem[Dewangan et~al.(2001)]{dewangan01}
Dewangan G.C., Singh K.P., Jones L.R., McHardy I.M., Mason K.O., 
Newsam A.M.,
2001, MNRAS, 325, 1616

\bibitem[Dwelly et~al.(2005)]{dwelly05}
Dwelly T., Page M.J., Loaring N.S., Mason K.O., McHardy I., Gunn K.,
Sasseen T., 2005, MNRAS, 360, 1426

\bibitem[Ferrarese \& Merritt(2001)]{ferrarese01}
Merritt D. \& Ferrarese L.,
2001, ApJ, 547, 140

\bibitem[Franceschini et~al.(2003)]{franceschini03}
Franceschini et~al., 2003, MNRAS, 343, 1181

\bibitem[Gallo et~al.(2004)]{gallo04}
Gallo L.C., Tanaka Y., Boller Th., Fabian A.C., Vaughan S., Brandt W.N.,
2004, MNRAS, 353, 1064

\bibitem[Giacconi et~al.(2002)]{giacconi02}
Giacconi R., et~al., 
2002, ApJS, 139, 369

\bibitem[Gilli, Risaliti \& Salvati(1999)]{gilli99}
Gilli R., Risaliti G., Salvati M.,
1999, A\&A, 347, 424

\bibitem[Guedel et~al.(2001)]{guedel01}
Guedel M., et~al., 2001, A\&A, 365, L336

\bibitem[Gunn et~al.(2001)]{gunn01}
Gunn K.F., et~al., 2001, MNRAS, 324, 305

\bibitem[Gunn et~al.(2003)]{gunn03}
Gunn K.F., et~al., 2003, AN, 324, 105

\bibitem[Hasinger et~al.(1998)]{hasinger98}
Hasinger G., Burg R., Giacconi R., Schmidt M., Trumper J. \& Zamorani G.,
1998, A\&A, 329, 482

\bibitem[Hasinger et~al.(2001)]{hasinger01}
Hasinger G., et~al., 
2001, A\&A, 365, L45

\bibitem[Hasinger et~al.(2005)]{hasinger05}
Hasinger G., Miyaji T., Schmidt M.,
2005, A\&A, in press, (astro-ph/0506118)

\bibitem[Helsdon \& Ponman(2000)]{helsdon00}
Helsdon S.F. \& Ponman T.J., 
2000, MNRAS, 315, 356

\bibitem[Kinkhabwala et~al.(2002)]{kinkhabwala02}
Kinkhabwala et~al., 2002, ApJ, 575, 732

\bibitem[Loaring et~al.(2005)]{loaring05}
Loaring N.S., et~al., 
2005, MNRAS, 362, 1371


\bibitem[Maccacaro et~al.(1988)]{maccacaro88}
Maccacaro T., Gioia I.M., Wolter A., Zamorani G., Stocke J.T., 
1988, ApJ, 326, 680

\bibitem[Mainieri et~al.(2002)]{mainieri02}
Mainieri V., et~al.,
2002, A\&A, 393, 425

\bibitem[Mainieri et~al.(2005)]{mainieri05}
Mainieri V., et~al., 
2005, A\&A, 437, 805

\bibitem[Maiolino et~al.(2001)]{maiolino01}
Maiolino R., Marconi A., Salvati M., Risaliti G., Severgnini P., Oliva
E., La Franca F., \& Vanzi L., 
2001, A\&A, 365, 28

\bibitem[Mateos et~al.(2005a)]{mateos05a}
Mateos S., et~al., 2005a, A\&A, 433, 855

\bibitem[Mateos et~al.(2005b)]{mateos05b}
Mateos S., et~al., 2005b, A\&A, 444, 79

\bibitem[McHardy et~al.(1998)]{mchardy98}
McHardy I.M., et~al., 
1998, MNRAS, 295, 641

\bibitem[McHardy et~al.(2003)]{mchardy03}
McHardy I.M., et~al., 2003, MNRAS, 342, 802

\bibitem[Mittaz et~al.(1999)]{mittaz99}
Mittaz J.P.D. et~al.,
1999, MNRAS, 308, 233

\bibitem[Moran, Lehnert \& Helfand(1999)]{moran99}
Moran E.C., Lehnert M.D., Helfand D.J.,
1999, ApJ, 526, 649

\bibitem[Page et~al.(1997)]{page97}
Page M.J., Mason K.O., McHardy I.M., Jones L.R., Carrera F.J.,
1997, MNRAS, 291, 324

\bibitem[Page(2001)]{page01b}
Page M.J.,
2001, MNRAS, 328, 925

\bibitem[Page, Mittaz \& Carrera(2001)]{page01a}
Page M.J., Mittaz J.P.D. \& Carrera F.J., 
2001, MNRAS, 325, 575

\bibitem[Page et~al.(2003)]{pageetal03}
Page M.J., et~al.,
2003, AN, 324, 101

\bibitem[Page, Davis \& Salvi(2003)]{pagedavissalvi03}
Page M.J., Davis S.W. \& Salvi N.J., 
2003, MNRAS, 343, 1241

\bibitem[Pappa et~al.(2001)]{pappa01}
Pappa A., Georgantopoulos I., Stewart G.C., Zezas A.L.,
2001, MNRAS, 326, 995

\bibitem[Sako et~al.(2001)]{sako01}
Sako M., Kahn S.M., Paerels F., Liedahl D.A., 2000, ApJ, 543, L115

\bibitem[Setti \& Woltjer(1989)]{setti89}
Setti G. \& Woltjer L.,
1989, A\&A, 224, L21

\bibitem[Severgnini et~al.(2003)]{severgnini03}
Severgnini P., et~al.,
2003, A\&A, 406, 483

\bibitem[Seymour, McHardy \& Gunn(2004)]{seymour04}
Seymour N., McHardy I.M., \& Gunn K.F.,
2004, MNRAS, 352, 131

\bibitem[Silverman et~al.(2005)]{silverman05}
Silverman J.D., et~al., 
2005, ApJ, 618, 123

\bibitem[Str\" uder et~al.(2001)]{struder01}
Str\" uder L., et~al., 2001, A\&A, 365, L18

\bibitem[Treister et~al.(2004)]{treister04}
Treister E., et~al., 
2004, ApJ, 616, 123

\bibitem[Trentham, Sampson \& Banerji(2005)]{trentham05}
Trentham N., Sampson L. \& Banerji M.,
2005, MNRAS, 357, 783

\bibitem[Turner et~al.(1997)]{turner97}
Turner T.J., George I.M., Nandra K., Mushotzky R.F.,
1997, ApJS, 113, 23

\bibitem[Turner et~al.(2001)]{turner01}
Turner M.J.L. et~al., 2001, A\&A, 365, L27

\end{thebibliography}
\end{document}